\documentclass[]{article}
\usepackage[margin=2 cm]{geometry}
\usepackage{comment}
\usepackage{graphicx}
\usepackage{amsmath,amssymb,extarrows,mathtools,graphicx,subfigure,setspace}
\usepackage{cite}
\usepackage{epsfig}
\usepackage[section]{placeins}
\usepackage{slashed}
\usepackage{color}
\usepackage{caption}
\usepackage{amsmath}
\usepackage{hyperref}
\makeatother

\newcommand{\be}{\begin{equation}}
\newcommand{\bea}{\begin{eqnarray}}
\newcommand{\eea}{\end{eqnarray}}
\newcommand{\ba}{\begin{array}}
\newcommand{\ea}{\end{array}}
\newcommand{\ee}{\end{equation}}
\newcommand{\bes}{\begin{equation*}}
\newcommand{\beas}{\begin{eqnarray*}}
\newcommand{\eeas}{\end{eqnarray*}}
\newcommand{\bas}{\begin{array*}}
\newcommand{\eas}{\end{array*}}
\newcommand{\ees}{\end{equation*}}

\numberwithin{equation}{section}

\begin{document}
\onehalfspacing
\vfill
\begin{titlepage}
\vspace{10mm}

\begin{center}

\vspace*{15mm}
\vspace*{1mm}
{\Large  Hyperscaling Violating Solution in Coupled Dilaton-Squared Curvature Gravity}
 \vspace*{1cm}
 
{ Mahdis Ghodrati}

\vspace*{8mm}
{\footnotemark University of Michigan, Randall laboratory of Physics, Ann Arbor, MI 48109-1040, USA}

E-mail: {ghodrati@umich.edu}
 \vspace*{2mm}

\vspace*{2cm}

\end{center}

\begin{abstract}

Considering a model with Coupled Dilaton-Squared Curvature terms and a dilatonic potential which replace the cosmological constant, we present an analytic hyperscaling violating solution which is the generalization of arXiv:1305.3279 for non zero $\theta$. We show that a specific coupling between dilaton and the squared curvature terms produces a simple logarithmic behavior for the dilaton. By imposing the null energy condition and stability constraints, we investigate the allowed regions for the parameters of this model. We then study the perturbations around  $\mathrm{AdS_4} $ in the UV and $\mathrm{\mathrm{AdS}_2\times \mathbb{R}^2} $ in the IR of the Einstein-Weyl gravity which can support an intermediate hyperscaling violating solution in order to resolve the $\mathrm{IR} $ singularity of the metric in 4d.

\end{abstract}

\end{titlepage}

\section{Introduction}

Lifshitz and hyperscaling violating geometries have been studied extensively as in the context of gauge/gravity duality \cite{Maldacena}, they can be used to study phases of matters in condensed matter physics such as non-Fermi liquids.
\\

Lifshitz metric 
\begin{eqnarray}
ds^2_{d+2}=-r^{2z}dt^2+\frac{dr^2}{r^{2}}+\sum\limits_{i=1}^d  r^2 d\vec{x_i}^2 ,
\end{eqnarray}
has the following scaling symmetry
\begin{eqnarray}
r \to  \lambda^z r, \ \ \ \ \ \  x_i \to \lambda x_i .
\end{eqnarray}
This metric can be derived from gravity coupled to massive gauge fields \cite{fixedpoints}, and they do not respect the relativistic scaling symmetry.  It is a generalization of AdS metric where for the special case of $z=1$ is ${\mathrm{AdS}_{d+2}}$ and when $z \to \infty$ becomes  ${\mathrm{Ads_2 \times \mathbb{R}^{d}}}$.
\\
\footnotetext{On leave of absence.}
Generalizing Lifshitz to hyperscaling violating metric for non-zero $\theta$ exponent,
\begin{eqnarray}
ds^2_{d+2}=r^{-\frac{2\theta}{d}}(-r^{2z}dt^2+\frac{dr^2}{r^{2}}+\sum\limits_{i=1}^d d\vec{x_i}^2) ,
\end{eqnarray}
changes the scaling symmetries to
\begin{eqnarray}
t \to \lambda^z t, \ \ \ \  r \to \lambda^{-1} r,\ \ \ \  x_i \to \lambda x_i,\ \ \ \  ds_{d+2} \to \lambda^{\frac{\theta}{d}} ds_{d+2}.
\end{eqnarray}
This metric is also not Lorentz invariant and can be derived by adding a dilaton field to the Einstein-Maxwell action with a dilatonic potential.

The Lifshitz geometry is a possible candidate for describing the behaviors of strange metals, holographically \cite{0912.1061}. Also in order to investigate a system with Fermi surface, we can consider hyperscaling violating geometries in a specific range of parameters for its gravitational dual \cite{Dong:2012se}.

Although these two space-times have constant and therefore finite scalar curvature invariants, they both have null singularity in the ${\mathrm{IR}}$ which makes the infrared region incomplete \cite{Knodel:2013fua}\cite {Horowitz}  \cite{Lei} \cite{Cremonini}\cite{Copsey} \cite{Kachru}\cite{Peet} \cite{Mann}. On the field theory side this might suggest that the solution cannot be trusted in the IR unless these singularities being resolved.
 \\
It has been suggested in  \cite{Knodel:2013fua}, that for the 4-dimensional Lifshitz metric in Einstein-Weyl gravity, one can construct numerically a flow from $\mathrm{AdS_4}$ in the UV to an intermediate Lifshitz region and then to $\mathrm{AdS_2 \times \mathbb{R}^2}$ in the IR. As $\mathrm{AdS}$ space-times is free from singularity, this solution can resolve the IR singularities. 
\\

In this paper we generalize the solutions in \cite{Knodel:2013fua} to non-zero $\theta$ and study the effects of squared curvature terms on the solutions of hyperscaling violating backgrounds. 
In section 2, we derive the analytical solution by coupling the higher derivative terms to the dilaton field by a $\mathrm{g(\phi)}$ function. Letting $\mathrm{\theta}=0$ in our results, the solution of  \cite{Knodel:2013fua} can be re-derived. We study how our solution is being renormalized in $z$ by the effects of higher derivative gravity.
In section 3, we study the regions of $d$ and $\theta$ which can lead to physical solutions by satisfying the imposed constraints, most importantly, null energy condition and stability of the solution. We study several different special cases in the parameter region and investigate constraints in various physical limits such as for the strange metals.
\\

In section 4,  we consider a four dimensional metric ansatz and then study the perturbations around $\mathrm{AdS_2 \times \mathbb{R}^2}$ in the IR and $\mathrm{AdS_4}$ in the UV which can support hyperscaling violating solution in the intermediate region. We investigate the allowed parameter space region for constructing numerical flow and then for some specific initial values, analytically estimate the cross overs from each region of the parameter space to the next one for the complete free from singularity solution. The last section is devoted to conclusions and discussions.
\section{General Solution}

The action which gives a hyperscaling violating solution corrected by squared curvature terms is 
\begin{flalign} \label{eq:action}
S=\int{d^{d+2}x\sqrt{-g}\bigg(R+V(\phi)-\frac{1}{2}(\partial \phi)^2-f(\phi)F_{\mu \nu}F^{\mu \nu}+g(\phi)(\alpha_1 R_{\mu \nu \rho \sigma}R^{\mu \nu \rho \sigma}+\alpha_2 R_{\mu \nu} R^{\mu \nu}+\alpha_3 R^2)\bigg)} .
\end{flalign}
As discussed in\cite{Shaghoulian:2013qia}, one way to derive a hyperscaling violating solution and fix the exponents is that the higher derivative terms should be coupled to the scalar field $\phi$, by multiplying these terms to a $g(\phi)$ function.\\

For deriving hyperscaling violating solution, the ansatz metric is
\begin{eqnarray}\label{eq:ansatz}
ds^2_{d+2}=r^{2\alpha}\left(-r^{2z}dt^2+\frac{dr^2}{r^{2}}+\Sigma \ r^2 d\vec{x_i}^2\right), \ \ \ i=1,2..,d.
\end{eqnarray}
We have set the AdS radius to one; $L=1$.
Here $\alpha=-\frac{\theta}{d}$, where $\theta$ and z are hyperscaling violation exponent and dynamical exponent respectively.\\
Taking the variation of the action and neglecting the surface terms, the Einstein's equations can be derived,

\begin{eqnarray}\label{Eineq}
E_{\mu \nu}=T_{\mu \nu} ,
\end{eqnarray}

where\footnote{We use the notation of \cite{Carroll} and $R_{[\mu |\beta | \nu] \alpha}=\frac{1}{2}(R_{\mu \beta  \nu \alpha}-R_{\nu \beta  \mu \alpha})$.}
\begin{eqnarray}\label{Emunu}
E_{\mu \nu}&=&R_{\mu \nu}-\frac{1}{2}g_{\mu \nu}R+g(\phi)\Big(-\frac{1}{2}g_{\mu \nu}\big(\alpha_1 R_{\alpha \beta \rho \lambda}R^{\alpha \beta \rho \lambda}+\alpha_2 R_{\alpha \beta}R^{\alpha \beta}+\alpha_3 R^2\big)+2\alpha_1 R_{\mu \alpha \beta \rho}R_\nu^{\alpha \beta \rho}+\nonumber\\
&&2\alpha_2 R_{\mu \alpha}R_\nu^\alpha+ 2\alpha_3 R R_{\mu \nu}\Big)+g_{\mu \nu}\big(\alpha_2 \nabla_\alpha \nabla_\beta g(\phi)R^{\alpha \beta}+2\alpha_3 \Box g(\phi)R\big)+4\alpha_1 \nabla^\alpha \nabla^\beta g(\phi)R_{[\mu |\beta | \nu] \alpha}\nonumber\\
&&+\alpha_2 \Big(-2\nabla_\alpha \nabla_\nu g(\phi)R_\mu^\alpha+\Box g(\phi)R_{\mu \nu}\Big)-2\alpha_3 \nabla_\mu \nabla_\nu g(\phi)R ,
\end{eqnarray}

and
\begin{eqnarray}\label{Tmunu}
T_{\mu \nu}&=&\frac{1}{2}\partial_\mu \phi \partial_\nu \phi +2 f(\phi)\left(F_{ \mu} ^\rho F_{ \nu \rho}-\frac{1}{4}g_{\mu \nu} F_{ \rho \sigma}F^{ \rho \sigma}\right)+\frac{1}{2}g_{\mu \nu}\left(V(\phi)-\frac{1}{2}\partial^\rho \phi \partial_\rho \phi \right).
\end{eqnarray}

These equations need to be supplemented with the Maxwell and scalar equations of motion,

\begin{flalign}\label{eq:max}
\nabla_\mu \left(f(\phi)F^{\mu \nu}\right)=0,
\end{flalign}
\begin{flalign}
\Box \phi-f'(\phi)F_{\mu \nu}F^{\mu \nu}+g'(\phi)\left(\alpha_1 R_{\mu \nu \rho \sigma}R^{\mu \nu \rho \sigma}+\alpha_2 R_{\mu \nu} R^{\mu \nu}+\alpha_3 R^2\right)=&-\frac{dV(\phi)}{d\phi},
\end{flalign}
where $V(\phi)=V_0 e^{\gamma \phi}, f(\phi)=e^{\lambda \phi}, g(\phi)=e^{\eta \phi}$ and prime denotes the derivative with respects to the argument.
\\
\\
Using the metric ansatz, the Maxwell equation \ref{eq:max} leads to
\begin{eqnarray}\label{maxsol}
F_{ rt}=\rho e^{-\lambda \phi}r^{(2-d)\alpha+z-d-1}.
\end{eqnarray}
For solving the Einstein's equations more easily, we combine the various components of the energy-momentum tensor in the following way

\begin{eqnarray}\label{Tcomp}
T_t^t-T_r^r&=&-\frac{1}{2} r^{2-2 \alpha } (\partial_r \phi)^2, \nonumber\\
T_x^x-T_t^t&=&2 \rho^2 r^{-2d (\alpha+1) } e^{-\lambda \phi} , \nonumber\\
T_r^r+T_x^x&=&V_0 e^{\gamma \phi}.
\end{eqnarray}
On the other hand, considering a logarithmic function for the dilaton as $\phi(r)=\phi_0+\beta \log r$ implies that $\eta=\frac{2\alpha}{\beta}$ and then one finds

\begin{eqnarray*}
\frac{r^{2 \alpha }(E_t^t-E_r^r)}{d(1+\alpha )}&=&1-z-\alpha+2e^{\eta  \phi _0}\Big[2\alpha _1\big(z(1-z)(1+d+\alpha (d-2))+z^3+\alpha+\alpha ^2(1-z-\alpha )-1\big)+\nonumber\\
&&(z+\alpha -1)\big(\alpha _2\left(d+z^2+(d+z)\alpha \right)+\alpha _3\left(\left(1+\alpha ^2\right)d(d+1)+2(d+z)(z+\alpha (d+1))\right)\big)\Big],
\end{eqnarray*}
\begin{eqnarray*}
\frac{r^{2\alpha }(E_x^x-E_t^t)}{(z-1)(z+d(1+\alpha ))}&=&1-2e^{\eta  \phi _0}\Big(2\alpha _1\left(1+z^2-\alpha ^2-z(d+(d-1)\alpha )\right)-\alpha _2\left(d+z^2+d \alpha +z \alpha \right)\\
&&-\alpha _3\left(2z^2+d(d+1)(1+\alpha )^2+2z(d+(d+1)\alpha )\right)\Big),
\end{eqnarray*}
\begin{eqnarray*}
r^{2\alpha}(E_r^r+E_x^x)&=&(d+z+d \alpha -1) (d+z+d \alpha )+e^{\eta  \phi _0} (1+\alpha )\times\\
&&\Big(2\big(d^2 (1+\alpha )^2 (3 \alpha -1)+d \left((\alpha-3)(\alpha^2-1)+2z(\alpha(4\alpha+3)-(2+(z-2)z))\right)\\
&&+2 z (1-\alpha +z (z+2 \alpha -1))\big) \alpha _1-\big(d z \left(2 z^2-2 (1+\alpha )^2-z (3+\alpha )\right)-d^3 (\alpha -1) (1+\alpha )^2\\
&&-2 z^2 (z+\alpha )+d^2 (1+\alpha ) \left(z (2+z)-2 z \alpha -\alpha ^2-3\right)\big) \alpha _2\\&&-\left(d^2 (1+\alpha )^2+2 z (z+\alpha )+d (1+\alpha ) (1+2 z+\alpha )\right) (d (d-3+2 z+(d-1) \alpha )-2 z) \alpha _3\Big).
\end{eqnarray*}

Now changing the basis of $\alpha$ corrections, we will write the solution in $\alpha_{GB}$, $\alpha_R$ and $\alpha_W$ basis which is related to the previous basis by
\begin{eqnarray}
\alpha_1&=&\alpha_{GB}+\alpha_W, \nonumber\\
\alpha_2&=&-4\alpha_{GB}-\frac{4}{d}\alpha_W, \nonumber\\
\alpha_3&=&\alpha_{GB}+\frac{2}{d(d+1)}\alpha_W+\alpha_R.
\end{eqnarray}
\\
In this new basis the general Lagrangian of the theory is \cite{Knodel:2013fua},
\begin{eqnarray}
\mathcal{L}=\alpha_{W} C_{\mu\nu\rho\sigma}C^{\mu\nu\rho\sigma}+\alpha_{GB} G+\alpha_{R}R^2 ,
\end{eqnarray}
where $C$ is the Weyl tensor and $G$ is the Gauss-Bonnet combination with the following definitions
\begin{eqnarray}
C_{\mu\nu\rho\sigma}=R_{\mu\nu\rho\sigma}-\frac{1}{d-1}(g_{\mu[\rho} R_{\sigma_]\nu}-g_{\nu[\rho}R_{\sigma]\mu})+\frac{1}{d(d-1)}g_{\mu[\rho}g_{\sigma]\nu}R,
\end{eqnarray}
and 
\begin{eqnarray}
G=R_{\mu\nu\rho\sigma}R^{\mu\nu\rho\sigma}-4R_{\mu\nu}R^{\mu\nu}+R^2.
\end{eqnarray}
By combining the above equations, considering $\alpha=-\theta/d$ and also using equation \ref{maxsol} and after some algebras, one find the solution in the new basis
\begin{eqnarray*}
ds^2&=&r^{-2\frac{\theta}{d}}\left(-r^{2z}dt^2+\frac{dr^2}{r^{2}}+r^2 d\vec{x}^2\right), \\
F_{rt}&=& \rho \hspace*{2mm}e^{\frac{2}{\beta}(d-\theta+\theta/d)\phi_0}r^{d+z-\theta-1} , \\
e^{\phi}&=&e^{\phi_0}r^\beta, 
\end{eqnarray*}

which the parameters of the solution, $\beta$, charge and potential are
\begin{eqnarray} \label{eq:beta}
\beta^2 &=&2(d-\theta)(z-1-\frac{\theta}{d})+\frac{4(d-\theta)}{d^3}e^{\eta \phi _0}\Big(\frac{2 z d^3(d-1) (z-1) (d-z-\theta +2)}{1+d}\alpha_W+\nonumber\\
&&d (d (1-z)-\theta ) \left(2z(d(d+z-\theta)-\theta)+(d+1)(\theta-d)^2\right)\alpha_R+\nonumber\\
&&(1-d) (d-\theta ) \left(d^2(d-2)(z-1)-d ((4+d) z-4) \theta +(2+d) \theta ^2\right)\alpha_{GB}\Big),
\end{eqnarray}
\begin{eqnarray}\label{eq:rho}
\rho^2e^{-\lambda \phi_0}&=&\frac{(z-1)(d+z-\theta)}{2}\Big(1+\frac{e^{\eta \phi _0}}{d^2 (d+1)}\big(2d(d+1)\left( 2z(\theta -d(d+z-\theta ))-(d+1)(\theta -d)^2\right)\alpha_R\nonumber\\
&&+4 d z\text{  }(1-d)(\theta +d (z+\theta -d-2))\alpha_W+2 \left(1-d^2\right) \left(d^2(d-2-2\theta)+(d+2) \theta ^2\right)\alpha_{GB}\big)\Big),
\end{eqnarray}
\begin{eqnarray} \label{eq:V}
V_0e^{\gamma\phi _0}&=&(d+z-\theta )(d+z-\theta -1)+\frac{(d-\theta)e^{\eta \phi_0}}{d^3}\Big(\frac{4z d^2(1-d)(z-1) (d (z-2)-z+\theta )}{d+1}\alpha_W\nonumber\\
&&+d (d (3-d)+(1-d)(2 z-\theta )) \left(\left((\theta -d)^2-2z \theta \right)(d+1)+2d z(d+ z)\right)\alpha_R\nonumber\\
&&+(1-d) \big( d^2(d-2) \left(d^2+2 (z-1) z+d (4z-3)\right)-d \big(d(d-2)(1+3 d)+2 \left(4 d^2+d-2\right) z\nonumber\\
&&+2 (2+d) z^2\big) \theta +d \left(5 d-6+3 d^2+4 (d+3) z\big)\theta^2-(2+d (d+5)) \theta ^3\right)\alpha_{GB}\Big),
\end{eqnarray}

\begin{eqnarray}
\gamma=-\eta=\frac{2\alpha}{\beta},\ \ \ \  \  \ \  \alpha=-\frac{\theta}{d}, \hspace*{1cm}\lambda=\frac{2}{\beta}(\theta-d-\frac{\theta}{d}).\ \ \ \ \ 
\end{eqnarray}

One can check that the scalar equation is being satisfied accordingly and do not imply any further relation.\\
Therefore the action \ref{eq:action}  with the ansatz \ref{eq:ansatz} admits hyperscaling violating solution with an electric gauge potential and a logarithmic scalar field in the form $\phi=\phi_0+\beta \  \mathrm{log} \ r$.\\

For the consistency check, we may notice that if we let $\theta=0$ and set the dimension $d^\prime=d-1$, these solution will exactly match with the Lifshitz solution found in  \cite{Knodel:2013fua} and for their case of $d=1$ and 2, the factor of Gauss-Bonnet combination does not contribute to the equation of motion as we expect from the theory of Gauss-Bonnet gravity. However for our general case that we coupled the dilaton with higher derivative terms, the theory is no longer the simple Gauss-Bonnet gravity and it would be no longer necessary that the factor of $\alpha_{GB}$ be zero in these specific dimensions and in our solution, in the general case of $\theta \ne 0$, this factor is not zero for $d=1,2$.\\

As it is obvious from the above solution, the scalar field $\phi$ runs logarithmically and cause the coupling functions  $f(\phi)=e^{\lambda\phi_0} r^{\lambda\beta}$, which couples the dilaton to the gauge field, and also $g(\phi)$ which couples the dilaton to the higher derivative terms, runs from weak coupling in the IR ($r \to 0$) to the strong coupling in the UV when $(r\to \infty)$.
\\

The function $g(\phi)$ coupled to the squared curvature terms, changes the usual hyperscaling violating solution in the Einstein gravity in a non trivial way. This term induces corrections of order $z^3$ and $z^4$ in the solution as one can check that $\beta^2$ which is zero at Lifshitz solution, now is being corrected by order of $z^3$, the electric charge $\rho^2$ is being renormalized by order of $z^4$ and the potential $V_0$ by order of $z^3$. Also the maximum order of $z$ inducing by Gauss-Bonnet factor is 2 in all the quantities and $\alpha_R$ and $\alpha_W$ can also induce corrections of order $z^3$ and $z^4$. \\

One should notice that we cannot choose any arbitrary $d$, $z$ and $\theta$ and end up with a physical solution. In \cite{Peet} three constraints of null energy condition, causality $(z > 1)$, and $0<d-\theta<d$ has been assumed to derive the range of physical regions for the parameters of the theory $\alpha_W$, $\alpha_{GB}$, $\alpha_R$. Here in the next section, we consider the specific cases of the solution by assuming different combinations of $\alpha_{GB}, \alpha_R$, $\alpha_W$  terms  and plot the ranges of $z$ and $\theta$ for each case coming from the conditions of $\beta^2\ge 0$, i.e. scalar field solution should not be oscillatory, and also $\rho^2 \ge 0$ (NEC) coming from only studying the gravity side. 

Other constraints that can be imposed on the region of parameters, is the stability of the thermodynamics. For the general hyperscaling violating metric ansatz, there should be the following relation $\frac{d-\theta}{z}>0$, to satisfy the positivity of specific heat \cite{Dong:2012se}. We don't consider this constraint here for plotting the physical region. Considering this constraint also plus the other two, for the full corrected gravity theory when all $\alpha$ corrections are present gives a smaller region of $\theta<d$ and $0<z<1$ which is both stable and physical. 
\section{Specific Cases of the Solution}

In the following sections, we investigate several special cases of the solution by letting the different combinations of higher derivative terms to be zero. Then assuming the remaining factors of each $\alpha_R$, $\alpha_W$ or $\alpha_{GB}$ get positive values, we will plot the allowed regions for the parameters $d$, $z$ and $\theta$ to compare their qualitative behaviors in these different limits. 

\subsection{Hyperscaling Violating Solution in the Einstein-Weyl gravity }

For the case of Einstein-Weyl gravity, $\alpha_R=\alpha_{GB}=0$, and then we can read the hyperscaling violating solution in this setup. 
So Weyl solution is as follows
\begin{flalign}\label{nec}
\beta^2&= 2 (d-\theta ) \left(z-1-\frac{\theta }{d}\right)+\text{$\alpha_W$}\frac{8 e^{\eta  \text{$\phi $0}}z\text{  }(z-1)(d-1)(d-\theta ) (d-z-\theta +2)}{d+1}. 
\end{flalign}

 Notice, in order to have a solution, the right hand side of [\ref{nec}]  must be positive and this gives a constraint similar to the NEC. If we set $\alpha_W=0$, the case with no higher derivative gravity, then $(\theta-d)(\theta-d z +d)\ge 0$, which is the NEC in pure hyperscaling violating solution.
 Also the electric charge and the constant of scalar potential is 
\begin{flalign}
\rho^2 e^{-\lambda \phi_0}  =\frac{1}{2} (z-1) (d+z-\theta ) \left(1+\text{$\alpha_W$} \frac{4 e^{\eta  \text{$\phi $0}} z(\theta +d (z+\theta -d-2))(1-d)}{d (d+1)}\right),
\end{flalign}

\begin{flalign}
V_0 e^{\gamma \phi_0} =& (z+d-\theta -1) (z+d-\theta )+\text{$\alpha_W$}\frac{4e^{\eta  \text{$\phi $0}}z (z-1)(z(d-1)-2d+\theta )(d-\theta )(1-d)  }{d (d+1)}.
\end{flalign}

Again letting $\theta=0$ and $d^\prime=d -1$ in here, these equations will match the results of  \cite{Knodel:2013fua}. So this solution is the generalization of Lifshitz to HSV for the Einstein-Weyl gravity. 

Now we would like to study the allowed regions of parameters. Generally three mentioned constraints should be satisfied to have a meaningful, physical solution. Also one needs to make sure that the choices for $\alpha_W$ doesn't blow up the potential. Here we just choose $\alpha_W=1$, so we don't need to worry about this case. However in the next section we should consider this condition as well.
 
\begin{figure}[]
\centering
\parbox{15cm}{
\centering
\includegraphics[scale=0.5]{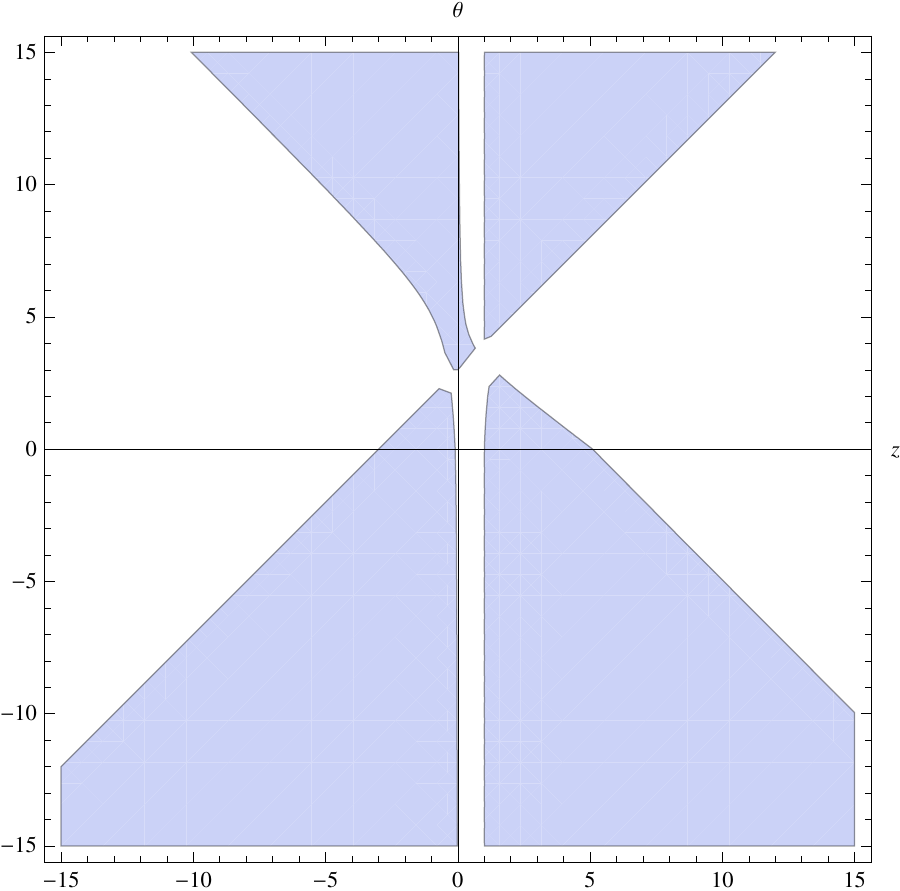}
\caption{Allowed region for  $z$, $\theta$ from both constraints of $\beta^2 \ge 0$ and $\rho^2 \ge 0$ for the Einstein-Weyl gravity, assuming $(\alpha_W=1,\ \phi_0=0,  d=4).$ \label{fig:weyl} }
\label{fig:2figsA4}}
\qquad
~ ~ ~ ~ ~ ~ ~ ~ ~ ~

\end{figure}

The allowed region from the constraint of having physical solution is being plotted in Figure \ref{fig:weyl}. The initial value for the dilaton has been assumed to be zero, also $\alpha_W=1$ and $d=4$.  One can notice specifically that the range of $0<z<1$ with general $\theta$ is not included in the physical space. 

Also if we let $d=1$, only the first term in $\beta^2$, $\rho^2$ and $V$ will remain; i.e. the factor of $\alpha_W$ is zero in all of them. So in this case we will reach to the pure hyperscaling violating solutions of previous works \cite{alishahiha:2012}. This means that the higher derivative corrections cannot induce any correction to hyperscaling violating and also Lifshitz space-times in three dimension, i.e $d+2=3$.

\subsection {Pure gravitational field with higher derivative terms $(\rho=0)$}
We now study the case where $\rho=0$  and matter field is being decoupled, so a pure gravitational field is being recovered. There are two ways that this can happen. One is $z=1$ which relates to a pure $\mathrm{AdS_{d+2}}$, as a pure $\mathrm{AdS}$ background describes bulk without matter field.
The second possibility is when:

\begin{flalign} \label {eq:alpha}
\alpha_W=& -\frac{e^{-\eta  \text{$\phi $0}}d (1+d) }{4 z(d-1)\left(d^2+2d-d z-\theta -d \theta \right)}.
\end{flalign}

If we let $\theta=0$ we will get the Lifshitz result\cite{Knodel:2013fua}
\begin{flalign} 
\alpha_W=-\frac{d+1}{4 z (d-1) (2+d-z)}.
\end{flalign}

So in this case $\rho^2=0$ and

\begin{flalign}
\beta^2=&-\frac{2 \theta (d+1) (d-\theta )^2  }{d \left(d^2-\theta -d (-2+z+\theta )\right)},
\end{flalign}

\begin{flalign}
V_{0} e^{\gamma \phi_0}= &(d+z-\theta -1) (d+z-\theta )-\frac{(z-1) (d-\theta ) (d (z-2)-z+\theta )}{\theta +d (z+\theta -d-2)} \ .
\end{flalign}

If we consider the case where $\theta=0$ corresponding to Lifshitz metric, and also consider  $d=d^\prime+1$ instead of $d=d^\prime+2$, we will get $\beta^2=0$. So the dilaton would be a constant value. Therefore equation (2.24) of  \cite{Knodel:2013fua}, corresponds to a pure gravitational Lifshitz solution in this limit. However for the hyperscaling violating space times which $\theta \ne 0$, for the limit of $\rho \to 0$, $\beta^2$ is no longer zero and therefore the dilaton would not be necessary a constant and we cannot recover purely gravitational solution unless $\theta=d$. It is because even without the charge, the potential of the dilaton can source the running of the dilaton and also breaks the scale invariance of the theory. This is one difference between Lifshitz and hyperscaling violating solution at this limit.  

For the conformal gravity where $\alpha_W \rightarrow \infty$, there would be two possible values for the dynamical exponent $z$ which are the the roots of the denominator of $\alpha_W$ in equation \ref{eq:alpha}. One of them is $z=0$, where in this case $\alpha_W \to \infty$, $\rho^2=0$ and 
\begin{flalign}
\beta^2=&-\frac{2 \theta }{d}\frac{(d+1) (d-\theta )^2 }{ d (d+2)- \theta (d+1)},
\end{flalign}
\begin{flalign}
V_{0} e^{\gamma \phi_0}= &\frac{(d+1) (d-\theta )^3}{d (d+2)-\theta (d+1) }.
\end{flalign}

The other case is $z=d+2-\theta -\frac{\theta }{d}$. At this $z$, generally both $V$ and $\beta^2$ will blow up. One case which can make both of them well behaved is when $d=\theta$, then $z=1$ and we will again end up with an AdS space-time and pure gravity theory.  The other case is when $\theta=0$ corresponding to Lifshitz background. Then from the condition on $\beta^2$, $z=d+2$ and from the condition on $V^2$, $d=2$. So for the case of $z=4,  d=2, \theta=0$ conformal gravity has a solution, consistent also with  \cite{Knodel:2013fua}. Therefore at this limit, hyperscaling violating background has not any further well behaved solution more than Lifshitz' solutions.     
\\

\begin{figure}[!ht] 
\centering
\parbox{17cm}{
\centering
\includegraphics[scale=0.5]{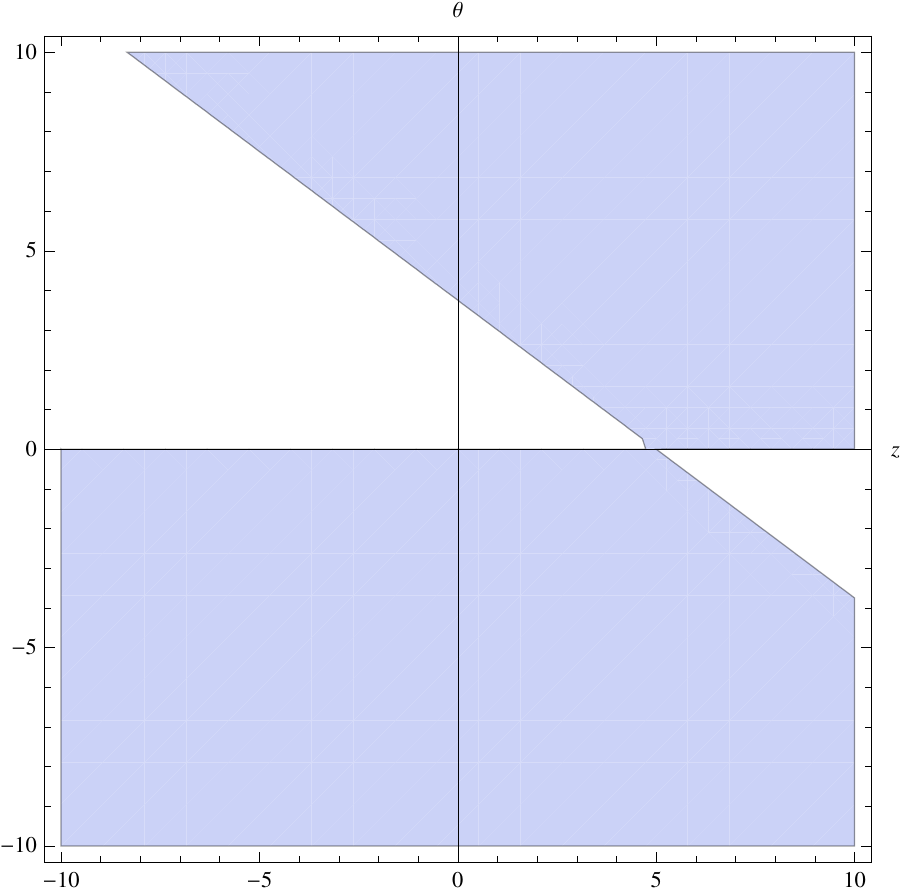}
\caption{Allowed region for  $z$, $\theta$ from the constraint of $\beta^2 \ge 0$, for the case of $\rho^2=0, d=4.$ \label{fig:rho}  }
\label{fig:2figsA5}}
\qquad

~ ~ ~ ~ ~ ~ ~ ~ ~ ~

\end{figure}

Again we would like to see which ranges will give us a physical solution in this case where $\rho^2 \to 0$. We should satisfy the constraint of $\beta^2 \ge 0$ and $d-\theta >0$ which corresponds to stability and also we need to make sure that the parameters in these regions do not blow up the potential $V(\phi)$. Figure~\ref{fig:rho} shows the allowed region for this limit. 
\\

\subsection{Hyperscaling Violating Solution in Gauss-Bonnet gravity }
If we assume $\alpha_W=0$ and $\alpha_R=0$, the solution in Gauss-Bonnet gravity would be 

\begin{flalign}
\beta^2=& 2 (d-\theta ) \left(z-\frac{\theta }{d}-1\right)+\text{$\alpha_{GB}$}\frac{4 e^{\eta  \text{$\phi $0}}  (1-d)(d-\theta )^2 \left((d-2) d^2 (z-1)-d ((4+d) z-4) \theta +(d+2) \theta ^2\right)}{d^3},
\end{flalign}

\begin{flalign}
\rho^2 e^{-\lambda \phi_0}=\frac{1}{2} (z-1) (d+z-\theta ) \left(1+ \text{$\alpha_{GB}$}\frac{e^{\eta  \text{$\phi $0}}2 \left(1-d^2\right)\left(d^2 (d-2 \theta -2)+(d+2) \theta ^2\right)}{d^2 (d+1)}\right),
\end{flalign}

\begin{flalign}
V_0 e^{\gamma \phi_0}&=(d+z-\theta -1) (d+z-\theta )+ \frac{e^{\eta  \text{$\phi $0}} \text{$\alpha_{GB}$}}{d^3}(1-d)\text{  }(d-\theta ) ((-2+d) d^2 (d^2+2 (-1+z) z+d (-3+4 z))\\&\nonumber  -d ((-2+d) d (1+3 d)+2 (-2+d+4 d^2) z+2 (2+d) z^2) \theta +d (-6+5 d+3 d^2+4 (3+d) z) \theta ^2-(2+d (5+d)) \theta ^3).
\end{flalign}
\begin{figure}[!ht]
\centering
\parbox{15cm}{
\centering
\includegraphics[scale=0.5]{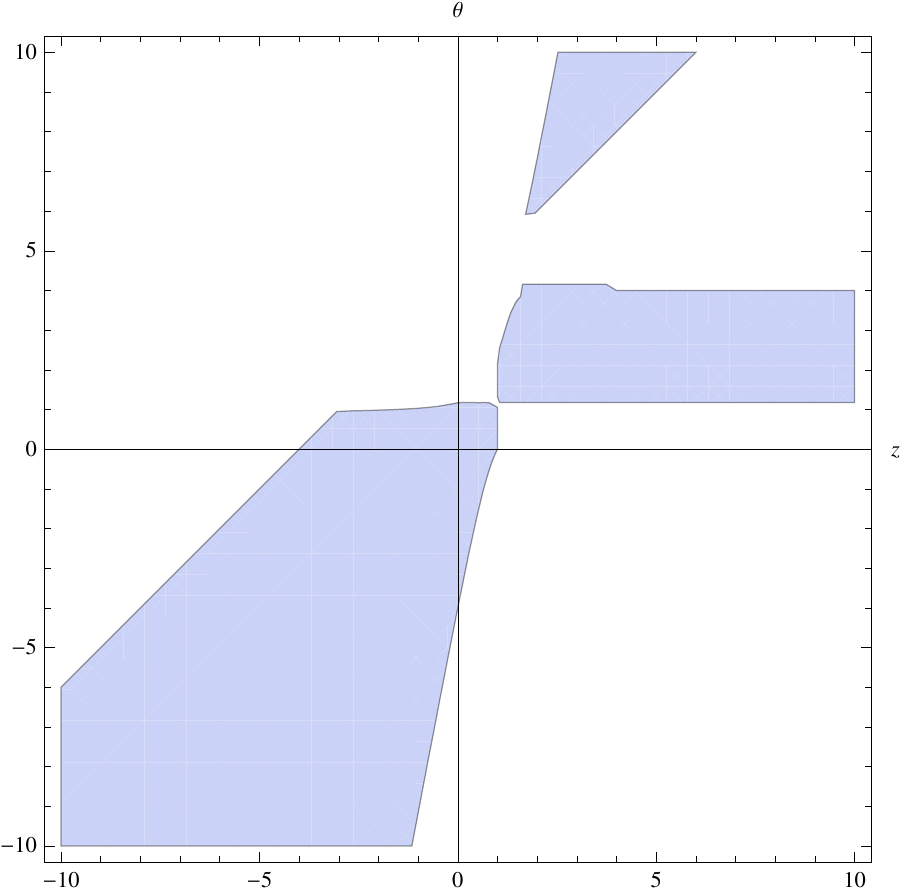}
\caption{Allowed region for  $z$, $\theta$ from the constraint of $\beta^2 \ge 0$, $d=4$ for Gauss-Bonnet case where $\alpha_W=\alpha_R=0 and \alpha_{GB}=1.$ \label{fig:gauss} }
\label{fig:2figsA2}}
\end{figure}
\\

As can be seen from Figure~\ref{fig:gauss} , ($d=4$), three distinct regions are allowable. Notably,  the region of $z>1$ with $ 1<\theta < d=4$ is present in the allowed region. Particular changes happen at $z=1$, $\theta=1$ and $\theta=d=4$. For Lifshitz solution where $\theta=0$, the region of $-d=-4<z<1$, is physical. Also for AdS case where $z=0$, $-d=-4<\theta<1$ is within the allowable ranges. It is worth to mention that changing the dimension, does not change these qualitative behaviors much.
\subsection{Hyperscaling Violating Solution in  $\mathrm{R^2}$  gravity }
We then study the solution for the case where $\alpha_W=\alpha_{GB}=0$. In this case the solution would be 

\begin{flalign}
\beta^2=&2 (d-\theta ) \left(z-\frac{\theta }{d}-1\right)+\text{$\alpha_R$}\frac{4}{d^2} e^{\eta  \text{$\phi $0}}\text{  }(d-\theta ) (d (1-z)-\theta ) \left(2 z (d (d+z-\theta )-\theta )+(1+d) (\theta -d)^2\right),
\end{flalign}

\begin{flalign}
\rho^2 e^{-\lambda \phi_0}=(z-1) (d+z-\theta ) \left(\frac{1}{2}+ \text{$\alpha_R$}\frac{e^{\eta  \text{$\phi $0}}}{d}\text{  }\left(-(1+d) (\theta -d)^2+2 z (-d (d+z-\theta )+\theta )\right)\right),
\end{flalign}

\begin{flalign}
V_0 e^{\gamma \phi_0}&=(d+z-\theta -1) (d+z-\theta )+\text{$\alpha_{R}$} \frac{e^{\eta  \text{$\phi $0}}}{d^2}((3-d) d+(1-d) (2 z-\theta )) (d-\theta ) \left(2 d z (d+z)+(1+d) \left(-2 z \theta +(\theta -d)^2\right)\right).
\end{flalign}

\begin{figure}[]
\centering
\parbox{15cm}{
\centering
\includegraphics[scale=0.5]{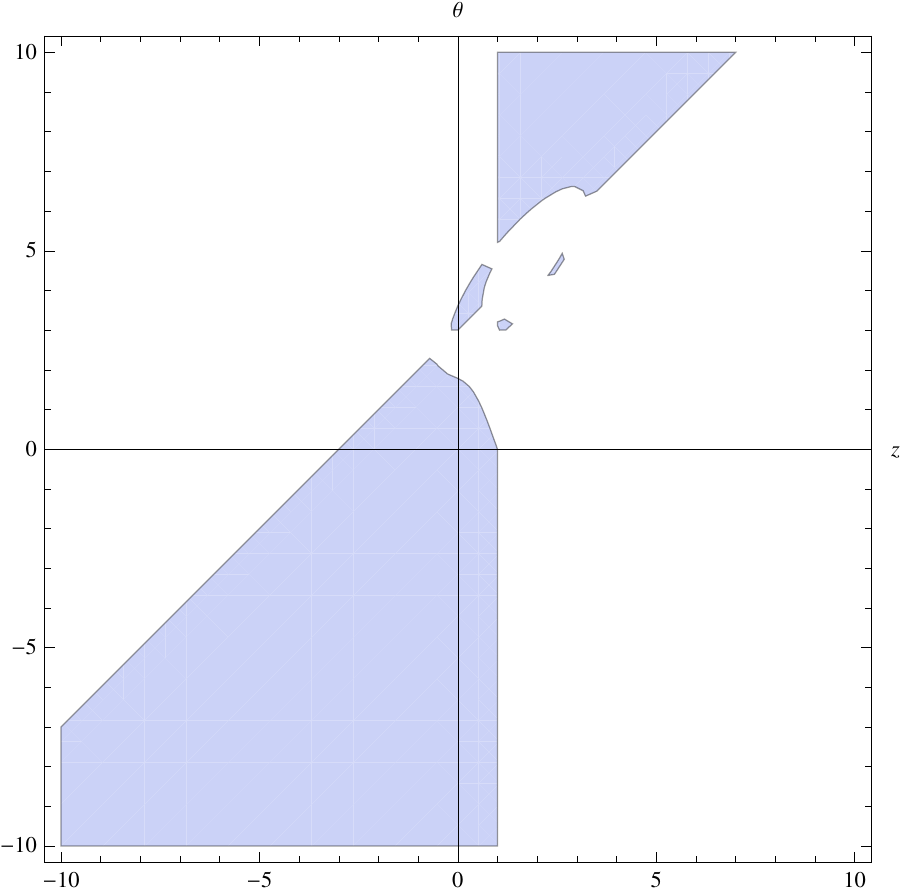}
\caption{Allowed region for  $z$, $\theta$ from both constraint of $\beta^2 \ge 0$ and $\rho^2 \ge 0$ for the $R^2$ gravity, assuming $(\alpha_R=1,\ \phi_0=0,  d=4).$  }
\label{fig:2figsA3}}
\qquad
\end{figure}
As can be noticed, the region of $z>0$, $1< \theta <4$ is not present anymore in $R^2$ theory.


\subsection{Hyperscaling Violating Solution in Gauss-Bonnet and $\mathrm{R^2}$  gravity }
In this section and the following sections, 3.6 and 3.7, we will turn off only one alpha correction and keep the other two to study the different combinations of $R^2$ corrections.\\

So for the case where $\alpha_W=0$, the solution is

\begin{flalign}
\beta^2=&2 (d-\theta ) (z-\frac{\theta }{d}-1)+\frac{4}{d^3} e^{\eta  \text{$\phi $0}} (d-\theta ) (\text{$\alpha _{GB}$}(1-d)(d-\theta )((d-2) d^2 (z-1)-d (-4+(d+4) z) \theta +(d+2) \theta ^2)+\nonumber \\& \frac{\text{$\alpha_ {R}$}}{d^2} (d (1-z)-\theta ) (2 z (d (d+z-\theta )-\theta )+(d+1) (\theta -d)^2)),
\end{flalign}

\begin{flalign}
\rho^2 e^{-\lambda \phi_0}=&\frac{1}{2} (z-1) (d+z-\theta ) \bigg(1+\frac{e^{\eta  \text{$\phi $0}}}{d^2 (1+d)} (2 \text{$\alpha_{GB}$} (1-d^2)(d^2 (d-2 \theta -2)+(d+2) \theta ^2)+\nonumber \\& 2 \text{$\alpha_{R}$}\text{  }d (d+1)(-(1+d) (\theta -d)^2+2 z (-d (d+z-\theta )+\theta )))\bigg),
\end{flalign}

\begin{flalign}
V_0 e^{\gamma \phi_0}&=(d+z-\theta -1) (d+z-\theta )+\frac{1}{d^3}e^{\eta  \text{$\phi $0}} (d-\theta ) (\text{$\alpha_{GB}$}(1-d)\text{  }((d-2) d^2 (d^2+2 (z-1) z+d (4 z-3))-\nonumber \\& d ((d-2) d (1+3 d)+2 (4 d^2+d-2) z+2 (d+2) z^2) \theta +d (3 d^2+4 (3+d) z+5d-6) \theta ^2-(2+d (5+d)) \theta ^3)+\nonumber \\& \text{$\alpha_R$} d(2 z-\theta +d (3-d-2 z+\theta ))(2 d z (d+z)+(1+d) ((d-\theta )^2-2 z \theta ))).
\end{flalign}
\begin{figure}[]
\centering
\parbox{15cm}{
\centering
\includegraphics[scale=0.5]{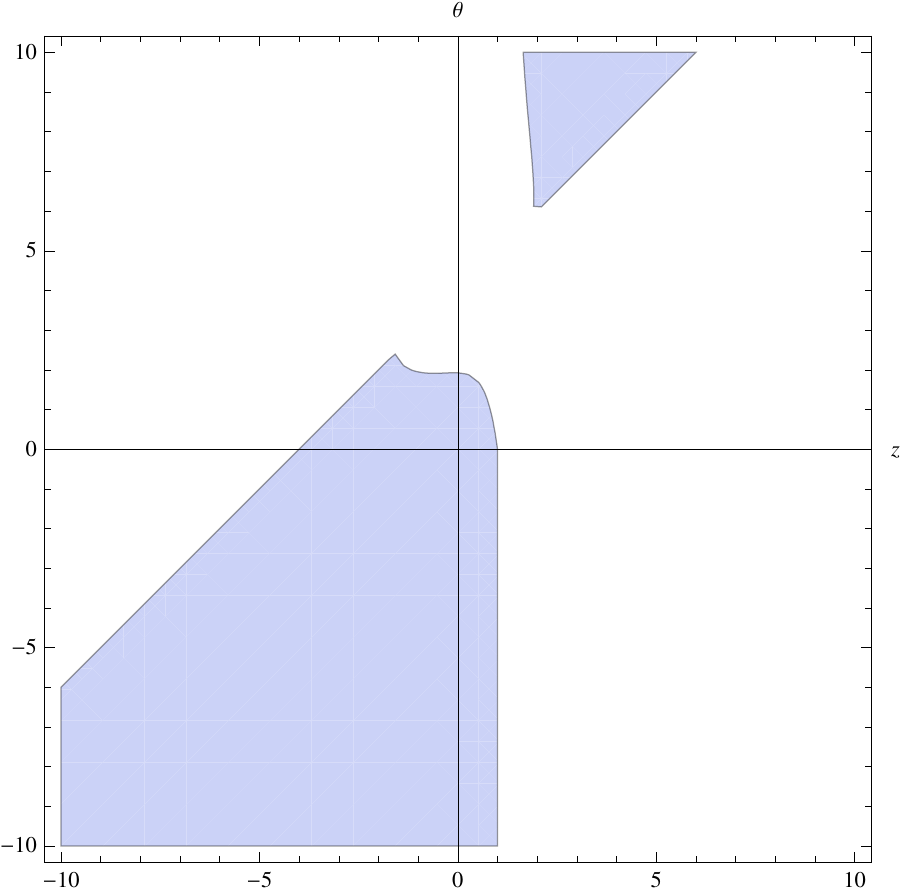}
\caption{Allowed region for  $z$, $\theta$ from both constraints of $\beta^2 \ge 0$ and $\rho^2 \ge 0$ for the $R^2$ and Gauss-Bonnet gravity, assuming $(\alpha_R=1,\alpha_{GB}=1,\alpha_W=0\ \phi_0=0,  d=4).$  }
\label{fig:2figsA6}}
\qquad
~ ~ ~ ~ ~ ~ ~ ~ ~ ~ 

\end{figure}

\subsection{Hyperscaling Violating Solution in Gauss-Bonnet and Weyl gravity }
We then study the solution for the case where $\alpha_R=0$. In this case the solution will be 

\begin{flalign}
\beta^2=&2 (d-\theta ) (z-\frac{\theta }{d}-1)+4 e^{\eta  \text{$\phi $0}} \frac{(d-\theta )}{d^3} (\text{$\alpha _W$}\frac{2 (d-1) d^3 (z-1) z\text{  }(d-z-\theta +2)}{1+d}+\nonumber \\&\text{$\alpha_{GB}$}(1-d)(d-\theta ) ((d-2) d^2 (z-1)-d (-4+(d+4) z) \theta +(d+2) \theta ^2)),
\end{flalign}

\begin{flalign}
\rho^2 e^{-\lambda \phi_0}=&(z-1) (d+z-\theta ) \left(\frac{1}{2}+e^{\eta  \text{$\phi $0}} (1-d)\left(\text{$\alpha_{GB}$} \frac{1 }{d^2 } \left(d^2 (d-2 \theta -2)+(d+2) \theta ^2\right)+\text{$\alpha_{W}$}\frac{2\text{  }}{d (1+d)}z (\theta +d (z+\theta -d-2))\right)\right),
\end{flalign}

\begin{flalign}
V_0 e^{\gamma \phi_0}&=(d+z-\theta -1) (d+z-\theta )+\frac{1}{d^3}e^{\eta  \text{$\phi $0}} (d-\theta ) (\text{$\alpha_{W}$}\frac{4 (1-d) d^2 (z-1) z\text{  }(d (z-2)-z+\theta )}{1+d}+\nonumber \\&\text{$\alpha_{GB}$}(1-d) ((d-2) d^2 (d^2+2 (z-1) z+d (4 z-3))-d ((d-2) d (1+3 d)+2 (-2+d+4 d^2) z+2 (2+d) z^2) \theta +\nonumber \\& d (-6+5 d+3 d^2+4 (3+d) z) \theta ^2-(2+d (5+d)) \theta ^3)).
\end{flalign}
\begin{figure}[]
\centering
\parbox{15cm}{
\centering
\includegraphics[scale=0.5]{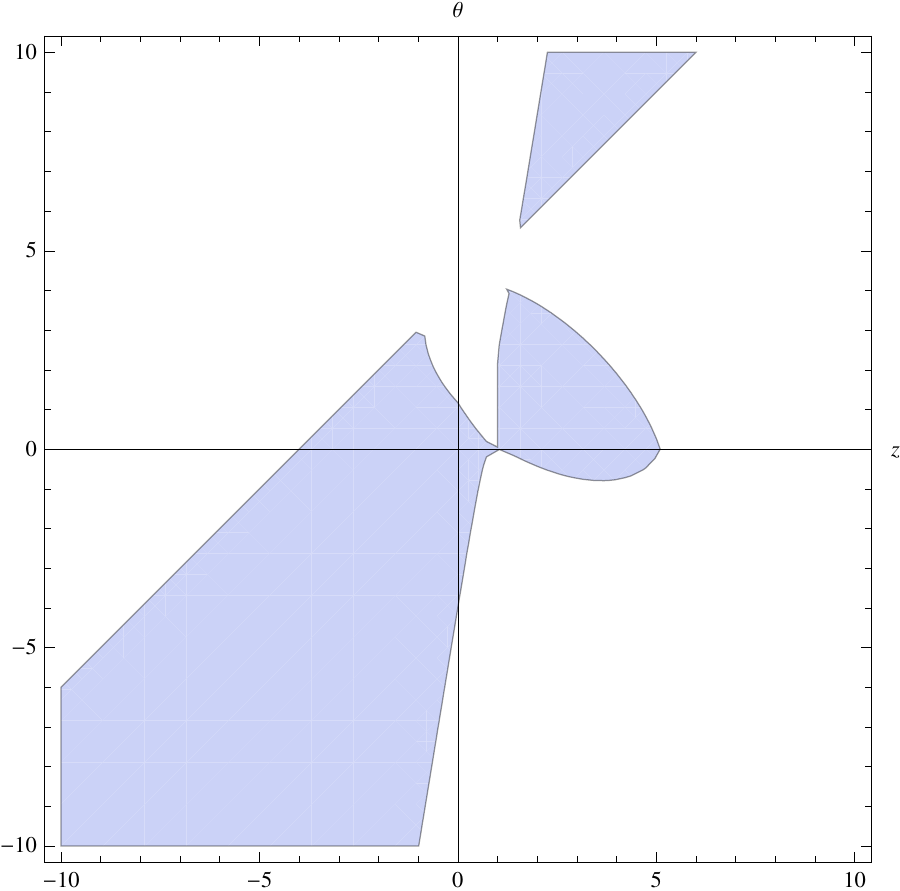}
\caption{Allowed region for  $z$, $\theta$ from both constraints of $\beta^2 \ge 0$ and $\rho^2 \ge 0$ for the  Gauss-Bonnet and Weyl gravity, assuming $(\alpha_W=\alpha_{GB}=1,\alpha_R=0\ \phi_0=0,  d=4).$  }
\label{fig:2figsA7}}
\qquad
~ ~ ~ ~ ~ ~ ~ ~ ~ ~ 

\end{figure}
One may notice how the solution from Lifshitz and hyperscaling violating from Einstein gravity, to Lifshitz solution in higher derivative corrected theory and then hyperscaling violating in higher derivative corrected theory   become more complicated as we eliminate the scaling symmetries and find solutions which can describe more exotic phases of matter by setting the parameter $\theta$. 

Also one may notice that in all solutions of hyperscaling violating a term of $d-\theta$ exist which can indicate again that the degrees of freedom of the theory effectively are in $d_{\mathrm{eff}}= d-\theta$ \cite{Sachdevbook}.   
 
\subsection{Hyperscaling Violating Solution in $R^2$ and Weyl gravity }
For the case where $\alpha_{GB}=0$, the solution will be 

\begin{flalign}
\beta^2=&2 (d-\theta ) (z-\frac{\theta }{d}-1)+4 e^{\eta  \text{$\phi $0}} (d-\theta ) ( \text{$\alpha_{W}$}\frac{2 (d-1) (z-1) z (2+d-z-\theta )}{1+d}\nonumber \\& +\frac{1 }{d^2}\text{$\alpha_R$} (d (1-z)-\theta ) (2 z (d (d+z-\theta )-\theta )+(d+1) (\theta -d)^2)),
\end{flalign}

\begin{flalign}
\rho^2 e^{-\lambda \phi_0}=&\frac{1}{2} (z-1) (d+z-\theta ) (1+e^{\eta  \text{$\phi $0}} (4 \text{$\alpha_W$} \frac{(1-d) }{d (1+d)} z (\theta +d (z+\theta -d-2))\nonumber \\&+2 \frac{\text{$\alpha_R$}}{d }(-(d+1) (\theta -d)^2+2 z (-d (d+z-\theta )+\theta )))),
\end{flalign}

\begin{flalign}
V_0 e^{\gamma \phi_0}&=(d+z-\theta -1) (d+z-\theta )+e^{\eta  \text{$\phi $0}} (d-\theta ) (\text{$\alpha $w} \frac{4 (1-d) (z-1) z (d (z-2)-z+\theta )}{(1+d)d}\nonumber \\&+\frac{\text{$\alpha_R$}}{d^2}\text{  }((3-d) d+(1-d) (2 z-\theta )) (2 d z (d+z)+(1+d) (-2 z \theta +(\theta -d)^2))).
\end{flalign}
\begin{figure}[ht!]
\centering
\parbox{15cm}{
\centering
\includegraphics[scale=0.5]{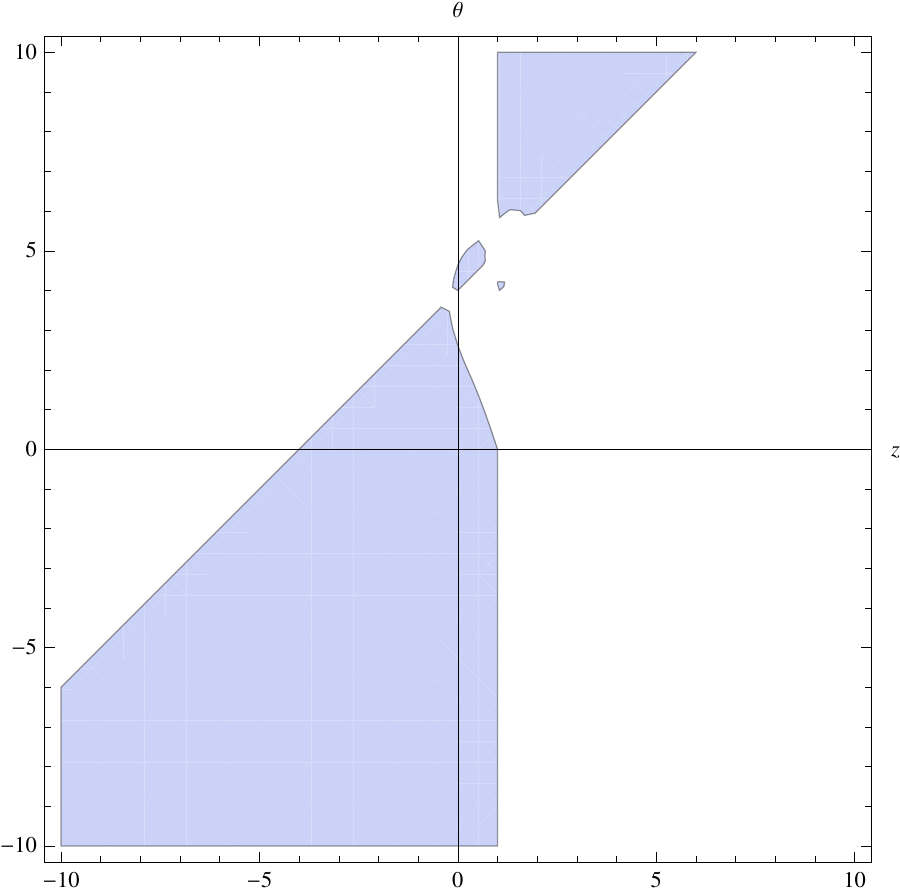}
\caption{Allowed region for  $z$, $\theta$ from both constraints of $\beta^2 \ge 0$ and $\rho^2 \ge 0$ for the  $R^2$ and Weyl gravity, assuming $(\alpha_W=\alpha_{R}=1,\alpha_{GB}=0\ \phi_0=0,  d=4). $  }
\label{fig:2figsA8}}
\qquad
~ ~ ~ ~ ~ ~ ~ ~ ~ ~

\end{figure}

\begin{figure}[]
\centering
\parbox{15cm}{
\centering
\includegraphics[scale=0.5]{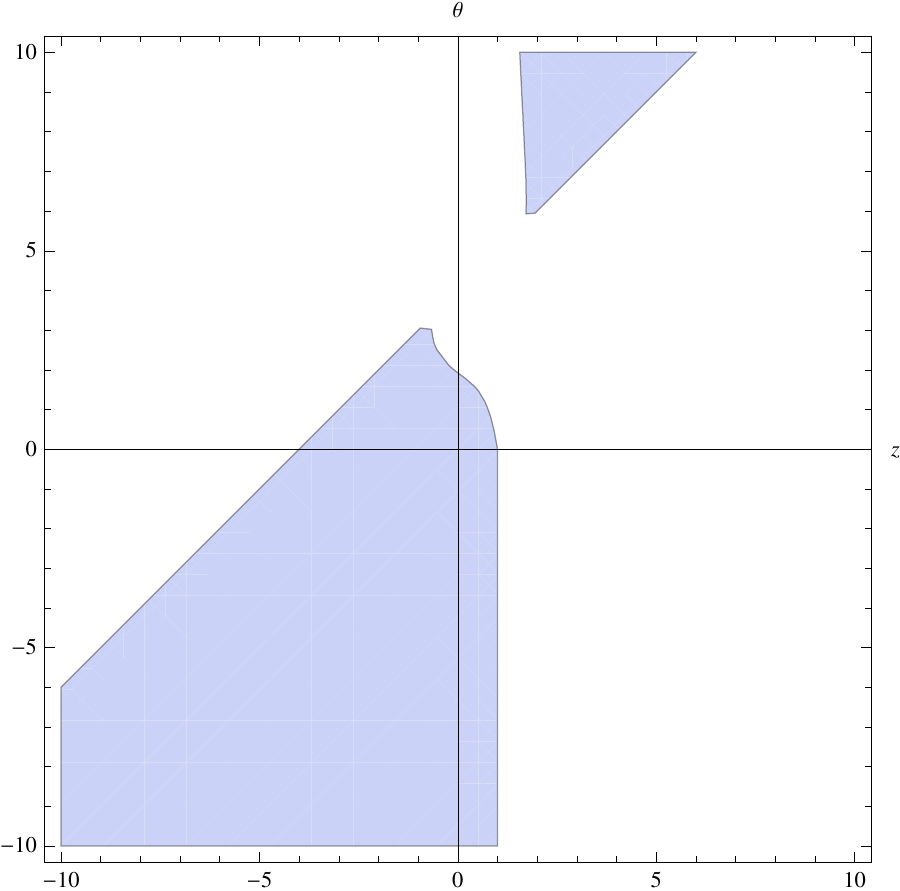}
\caption{Allowed region for  $z$, $\theta$ from both constraints of $\beta^2 \ge 0$ and $\rho^2 \ge 0$ for the most general higher derivative gravity theory, assuming $(\alpha_W=\alpha_{GB}=\alpha_{R}=1, \phi_0=0,  d=4).$  }
\label{fig:2figsA9}}
\qquad
~ ~ ~ ~ ~ ~ ~ ~ ~ ~

\end{figure}

We can compare the physical region of $\theta$ and $z$ for each combination of $\alpha$ corrections. For instance one can notice that Einstein-Weyl gravity is the least strict theory while $R^2$, would be the most strict theory on the physical region. The Gauss-Bonnet gravity has specifically an approximate region of  $z>1$ and $1< \theta <d$ in its physical solution which is not present in the $R^2$ gravity, and Einstein-Weyl theory can consist two more regions which placed in the second and forth quarter of space coordinate which are not present in the other two theories. 

As discussed in \cite{Dong:2012se}, the point of $z=1$, $\theta=-\frac{1}{3}$, is also special as it corresponds to $N$ D2-branes, in type IIA supergravity. It can be seen that this point is not within the physical region for $\alpha_{GB}$ gravity corrected term, Fig~\ref{fig:gauss}  and $\alpha_{GB}+\alpha_W$ terms, Fig~\ref{fig:2figsA9}.

If we consider also the constraint of $0<d-\theta <d$, corresponding to stability of the solution, for $d=4$, all the regions of $\theta>d$ would be eliminated from the plots. However we kept the thermodynamically unstable regions within the physical regions for the higher derivative corrected solution here. 

Other special limit is when $\theta=d$, where there is no region that the r.h.s of $\beta^2$ is positive in the squared curvature corrected gravity and so there is no region that dilaton is non-oscillatory in this limit.

Also one can notice that changing the dimension of the theory, $d$, does not change the general qualitative behavior much, and in each $d$ the allowed region in each theory approximately behave the same.  
One may also fix $\theta$, $z$ and $d$ and plot the regions for $\alpha_W$,$\alpha_{GB}$ and $\alpha_R$ as has been done in \cite{Peet}.

\subsection{The allowed regions for Strange Metals}

One special case where the degrees of freedom effectively live in one dimension, where $d-\theta=1$, corresponds to strange metals, a form of non-fermi liquid matter which their resistivities increase linearly by $T$ rather than by $T^2$ \cite{Sachdevbook}  \cite {Senthil} \cite{Musso} \cite{Sachdev2011de}. High temperature superconductors is one example of such materials. These metals have been studied using hyperscaling violating backgrounds in the context of holography \cite{Sachdev2014de}\cite{Roy}. \\

\begin{figure}[!ht]
\centering
\parbox{14cm}{
\centering
\includegraphics[scale=0.5]{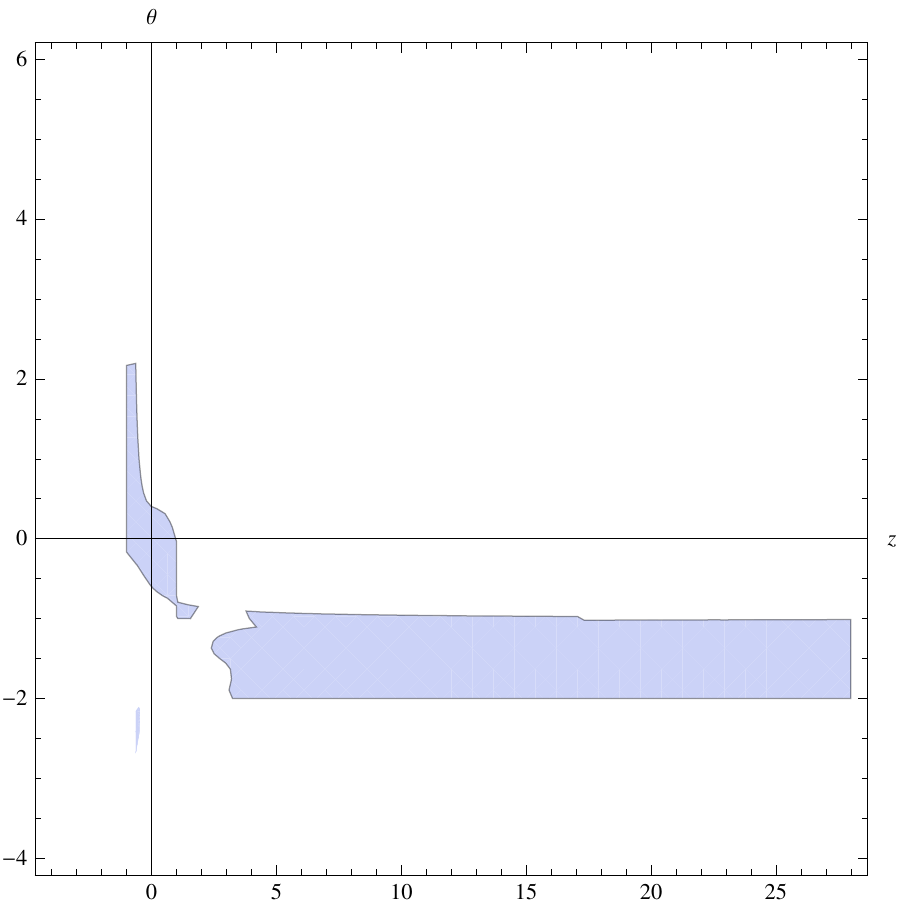}
\caption{Allowed region for strange metals where $d=\theta+1$ and all $\alpha$ corrections are on, $\alpha_R=\alpha_{GB}=\alpha_{W}=1$. \label{fig:metal} }
\label{fig:2figsA2}}
\end{figure}

Without higher derivative corrections, from NEC,  the relation of $z \ge 2-\frac{1}{\theta+1}$ gives the allowed region for the strange metals.  \cite{Dong:2012se}. Considering higher gravity corrections,  when all the $\alpha$ corrections are turned on, gives more complicated relations for our solution where the the allowed region of parameters is shown in Figure~\ref{fig:metal}.  One can see that the range of $-2< \theta <-1$ and $z  \ge 3$ is within the allowable region in the full higher derivative corrected theory, in agreement with the results without higher derivative gravity. 

However when only $\alpha_{GB}$ is on, all the region of $z>1$ with any $\theta$ is acceptable and so it is the least strict theory with biggest allowable region. On the other hand, $\alpha_R$ is the most strict theory as we also saw in the previous section, and only approximately a small region of $-2< \theta <4$, $-1<z <1$ would be allowable.

\section{Resolving the Singularity}
In this section we consider only a Weyl correction to the action coupled to the dilaton and investigate the singularity resolution. By choosing an appropriate $g(\phi)$ function, this term can lead to corrections to the effective potential in the deep IR which stabilize the dilaton at a finite value $\phi_I$ and the geometry would be free from singularity in this regime. In the UV limit of the theory as Weyl tensor vanishes, this correction would have no effect. \\ 

So the action that we consider is
\begin{eqnarray}
S=\int{d^{4}x\sqrt{-g}(R+V(\phi)-\frac{1}{2}(\partial \phi)^2-f(\phi)F_{\mu \nu}F^{\mu \nu}+g(\phi)C_{\mu \nu \rho \sigma}C^{\mu \nu \rho \sigma})},
\end{eqnarray}

where $g(\phi)=\frac{3}{4}\left(c_0e^{\eta \phi}+c_1\right)$ and $\alpha_W=\frac{3}{4}c_0$. 
The metric ansatz for constructing the flow is
\begin{eqnarray}\label{allsol}
ds^2&=&a^2(r)\left(-dt^2+dr^2+b^2(r)\left(dx^2+dy^2\right)\right).
\end{eqnarray}

We would call this new form of ansatz, the (a-b) gauge. The field parametrization in this new gauge is
\begin{eqnarray}
F_{rt}&=&\frac{Q}{f(\phi)b^2},
\end{eqnarray}

and the Einstein equations are

\begin{eqnarray}\label{e1}
E_{rr}&=&\frac{3 a'^2}{a^2}+\frac{b'^2}{b^2}+\frac{4 a' b'}{a b}+\frac{4g}{3a^2}\left(\frac{ b'^4}{2b^4}+\frac{ b''^2}{2b^2}-\frac{b'b^{(3)}}{ b^2}+\frac{ b'g' \phi '}{ g b^2}\left(\frac{ b'^2}{ b}-b''\right)\right),\nonumber\\
E_{tt}&=&-2\left(\frac{a''}{a}+\frac{b''}{b}\right)+\frac{a'^2}{a^2}-\frac{b'^2}{b^2}-\frac{4a'b'}{ab}-\frac{4g}{3a^2}\times\nonumber\\
&&\left(\frac{(b'b^{(3)})'}{b^2}-\frac{2b'^2b''}{b^3}-\frac{b''^2}{2b^2}+\frac{b'^4}{2b^4}+\frac{g'\phi''+\phi '^2 g''}{gb}\left(b''-\frac{b'^2}{b}\right)+\frac{g' \phi '}{gb}\left(2b^{(3)}-\frac{b'^3}{b^2}-\frac{b' b''}{b}\right)\right),\nonumber\\
\frac{E_{ii}}{b^2}&=&\frac{2 a''}{a}+\frac{b''}{b}-\frac{a'^2}{a^2}+\frac{2 a' b'}{a b}-\frac{4g}{3a^2}\times\nonumber\\
&&\left(\frac{ b^{(4)}}{2b}+\frac{ b'^4}{2b^4}-\frac{ b'^2 b''}{ b^3}+\frac{ g' \phi '}{g b}\left( b^{(3)}-\frac{ b'b''}{ b}\right)+\frac{1}{2g b}\left(b''- \frac{b'^2 }{b}\right)\left(\phi '^2g''+ g'\phi ''\right)\right),
\end{eqnarray}

with the right side
\begin{eqnarray}\label{e2}
T_{rr}&=&-\frac{Q^2}{b^4 a^2f }+\frac{1}{4} \phi '^2+\frac{a^2}{2} V, \nonumber\\
T_{tt}&=&\frac{Q^2}{b^4 a^2f }+\frac{1}{4} \phi '^2-\frac{a^2}{2} V, \nonumber\\
\frac{T_{ii}}{b^2}&=&\frac{Q^2 }{b^4 a^2f } -\frac{1}{4}\text{  }\phi '^2+\frac{a^2}{2} V, 
\end{eqnarray}

and the scalar equation is
\begin{eqnarray*}
\phi ''+2\left(\frac{a'}{a}+\frac{b'}{b}\right)\phi '+\frac{2 Q^2 f'}{b^4 a^2 f^2}+\frac{4g'}{3b^2a^2}\left(\frac{b'^4}{b^2}-\frac{2 b'^2 b''}{ b}+ b''^2\right)+a^2V'=0.
\end{eqnarray*}

As we can see from this equation, the derivative of the function $g(\phi)$ coupled to the metric function $b(r)$, affects the effective potential of the dilation.\\

Now we consider an initial solution of $\mathrm{AdS}_2\times \mathbb{R}^2$ in the $\mathrm{IR}$
\begin{eqnarray*}
a(r)=\frac{1}{r},\hspace*{1cm}b(r)=b_Ir,\hspace*{1cm}\phi(r)=\phi_I.
\end{eqnarray*}

Then from \eqref{e1} and \eqref{e2} we can derive
\begin{eqnarray*}
V(\phi_I)=1,\hspace*{1cm}\frac{Q^2}{b_I^4}=\frac{f(\phi_I)}{2}\left(1-\frac{4}{3}g(\phi_I)\right),
\end{eqnarray*}

and the scalar equation becomes
\begin{eqnarray*}
\frac{f'(\phi_I)}{f(\phi_I)}\left(1-\frac{4}{3}g(\phi_I)\right)+\frac{4}{3}g'(\phi_I)+V'(\phi_I)=0.
\end{eqnarray*}

Now by considering $f(\phi)=e^{\lambda \phi}, V(\phi)=V_0e^{\gamma \phi}$, we have the following solution in the IR
\begin{eqnarray*}
V_0 e^{\gamma_1 \Phi_I}&=&1,\nonumber\\
\frac{Q^2}{b_I^4}&=&\frac{e^{\lambda \Phi_I}}{2}\frac{(c_1-1)\eta-\gamma}{\lambda-\eta}, \nonumber\\
\Phi_I&=&\frac{1}{\eta}\log \left(\frac{\lambda(1-c_1)+\gamma}{c_0(\lambda-\eta)}\right).
\end{eqnarray*}

\subsection{IR Perturbations}
We now consider perturbations around the above IR, $\mathrm{AdS}_2 \times \mathbb{R}^2$ solution
\begin{eqnarray*}
a(r)=\frac{1}{r}+\delta a(r),\hspace*{1cm}b(r)=b_Ir+\delta b(r),\hspace*{1cm}\phi(r)=\phi_I+\delta \phi(r).
\end{eqnarray*}

For simplicity we can take $b_I=1$. Then from the Einstein and dilaton equations, there would be four coupled perturbative equations with maximum order of 4. 
\begin{flalign}
rr: r^2 \delta b^{(3)}+\frac{3 }{2g_I}\left(r^2 \delta a\right)'-2 \delta b'\frac{f_I}{f_I'g_I}\left(g_I'+\frac{3}{4}V_I'\right)-\frac{g_I'}{g_I}(r\delta \phi)'&= 0,
\end{flalign}
\begin{flalign}
tt:\left(r \delta  b^{(3)}\right)'+\frac{3}{2 g_I}\frac{\left( r^3 \delta  a'\right)'}{r^2}-\frac{g_I'}{g_I}\frac{(r \delta  \phi ')'}{r}+\frac{ g_I'}{g_I}\frac{\delta  \phi }{r^2}-2\frac{f_I}{f_I'g_I}\left(g_I'+\frac{3}{4}V_I'\right)\left(\frac{\delta  b'}{r}\right)'&=0, 
\end{flalign}
\begin{flalign}
ii: r^2 g_I \delta  b^{(4)}-3\left(r^2 \delta  a'\right)'-\frac{6 \delta  b}{r^2}-r g_I' \delta  \phi ''+2\frac{\delta  \phi }{r}\left(g_I'+\frac{3 V_I'}{4}\right)-2r^2\left(\frac{3}{4}+g_I\right)\left(\frac{\delta  b'}{r^2}\right)'&=0, 
\end{flalign}
\begin{flalign}
scalar: \delta  \phi ''-\frac{8}{3}r g_I'\left(\frac{ \delta  b'}{r^2}\right)'+\frac{4}{r}\left(\delta  a+\frac{\delta  b}{r^2} \right)V_I'+\frac{\delta  \phi }{r^2}\left(\frac{2f_I'}{ f_I}\left(\frac{4}{3}g_I'+V_I'\right)+\frac{ f_I''}{ f_I}-\frac{4 g_I f_I''}{3f_I}+\frac{4g_I''}{3 }+ V_I''\right)&=0.
\end{flalign}
One can read $\delta b^ {(4)} $ from ($ii$) and replace it in $(tt)$ and therefore end up with 3 equations with maximum order of 3.

Now considering the IR perturbations
\begin{eqnarray*}
\delta a(r)\sim r^{\nu-1},\hspace*{1cm}\delta b(r) \sim r^{\nu+1},\hspace*{1cm}\delta \phi(r)\sim r^{\nu},
\end{eqnarray*}

we can derive the following solutions for $\nu$
\begin{eqnarray*}
\nu_{1,2}&=&-1,\hspace*{2cm} \nu_3=2,\nonumber\\
\nu_{4,..,7}&=&\frac{1}{2}\pm\frac{1}{2}\left[1-\frac{2}{ 3\left( \gamma - c_1 \eta + \lambda _1\right)}\left[u_0\pm\left(24 \left(\gamma +\eta -c_1 \eta \right) \left(\gamma +\lambda -c_1 \eta \right) \left(c_1 \eta  \lambda -\left(\lambda +\gamma \right) \left(\eta +\gamma \right)\right)+u_0^2\right)^{\frac{1}{2}}\right]\right]^{\frac{1}{2}}, \nonumber\\
u_0&=&\left(\gamma +\left(1-c_1\right) \eta \right) \left(-2+3 \left(\gamma +\lambda \right)^2\right)+3c_1 \eta ^2\left(\gamma +\left(1-c_1\right) \lambda \right).
\end{eqnarray*}

\subsection{ Hyperscaling Violating Solution in the (a-b) gauge} \label{ab}
Now we need to write the hyperscaling violating solution that we have found in section 2 in a new gauge as \eqref{allsol} in order to match all three regions in a similar form of a metric.
Changing variable in \eqref{allsol}, in the way that $z\tilde{r}=r^{-z}, \tilde{x}=z^{1-\frac{1}{z}}x$, leads to
\begin{eqnarray}
ds^2=\tilde{L}^2\tilde{r}^{\frac{\theta}{z}-2}\left(-dt^2+d\tilde{r}^2+\tilde{r}^{2-\frac{2}{z}}d\tilde{x}^2\right),
\end{eqnarray}
where $\tilde{L}^2=L^2z^{\frac{\theta}{z}-2}$.\\

The hyperscaling violating solution in this gauge is
\begin{eqnarray}\label{eq:newsolg}
ds^2&=&\tilde{L}^2 r^{(1-\tilde{z})\theta-2}\left(-dt^2+dr^2+r^{2\tilde{z}}dx^2\right), \nonumber\\
\phi(r)&=&\phi_0+ (\tilde{z}-1)\beta \log r ,   \nonumber\\
F_{rt}&=& \rho e^{-\lambda \phi_0}r^{2(\tilde{z}-2)-(\tilde{z}-1)\theta} , \nonumber\\
\beta^2&=&\frac{(\theta-2)(2\tilde{z}+(\tilde{z}-1)\theta)}{\tilde{z}-1}+\frac{2c_0e^{\eta \phi_0}}{\tilde{L}^2}\frac{\tilde{z}(\theta-2)(3-4\tilde{z}+(\tilde{z}-1)\theta)}{\tilde{z}-1} , \nonumber\\
\rho^2 e^{-\lambda \phi_0}&=&\frac{\tilde{L}^2}{2}\tilde{z}(3-2\tilde{z}+(\tilde{z}-1)\theta)+\frac{c_0e^{\eta \phi_0}}{4}\tilde{z}(3-2\tilde{z}+(\tilde{z}-1)\theta)(6-8\tilde{z}+3(\tilde{z}-1)\theta) , \nonumber\\
V_0 e^{\gamma \phi_0}&=&\frac{(3-2\tilde{z}+(\tilde{z}-1)\theta)(2-\tilde{z}+(\tilde{z}-1)\theta)}{\tilde{L}^2}+\frac{c_0e^{\eta \phi_0}}{2\tilde{L}^4}\tilde{z}(\tilde{z}-1)(\theta-2)(3-4\tilde{z}+(\tilde{z}-1)\theta),
\end{eqnarray}
where $\eta=-\gamma=\frac{-\theta}{\beta}, \lambda \beta=\theta-4$ and $z=\frac{1}{1-\tilde{z}}$.

\section{Allowed Regions for the Numerical Solution}
To search for the numerical solution, we should consider some constraints on our extensive parameters.
In order to have an acceptable solution in the IR, from the terms for $Q^2$ and $\phi_I$, we get two constraints between $\lambda$, $\eta$, $\gamma$, $c_0$ and $c_1$ which are:
$\mathrm{\frac{ \left(\gamma +\eta -\eta  c_1\right) }{(\eta -\lambda )}}>0$ and $\mathrm{\frac{\lambda(1-c_1) +\gamma }{c_0(\lambda-\eta)} }>0$.  Now if we consider two cases of $\mathrm{c_1<1+\frac{\gamma}{\lambda}  }$ or $\mathrm{c_1>1+\frac{\gamma}{\lambda}  }$, for both case one can demonstrate that $\mathrm{c_0<0 }$. Therefore we will consider a negative value for $c_0$.\\

Then we would like to find all the acceptable regions for $\lambda$, $\eta$ and $c_1$. The conditions that we will impose, similar to  \cite{Knodel:2013fua} are:
 \\ \\ 
 1) $\mathrm{\lambda, \eta, \gamma >0}$. \\ \\
 2) There should be a region where $V_{\mathrm{eff}}(\phi_0)=0$. So the argument of the logarithm in the equation for $\phi_0$ should be positive.  \\ \\
3) $\mathrm{g(\phi_0) >0}$ which means $\mathrm{\frac{3}{4} (c_0 E^{\eta\phi_0}+c_1)>0}.$\\ \\
 4) $\mathrm{Q^2 >0}$, therefore $\mathrm{\frac{ \left(\gamma +\eta -\eta  c_1\right) }{(\eta -\lambda )}}>0.$\\ \\
 5) The perturbations should not be oscillatory, which means all the $\nu$'s that we have found should be real parameters, which put constraints on the terms which are under the radical.\\ \\ 
6) At least one of the $\nu$ should be negative and therefore one of the dilaton perturbation should be irrelevant.\\ 

Using these conditions we can specify different regions of the parameter space in the following figures. Similar to  \cite{Knodel:2013fua} which separated the parameters regions for the Lifshitz metric, we will do the same for the hyperscaling violating metric. So the green region is the allowed region, red is when $g(\phi)<0$, yellow is when $g(\phi)>0$, but $\nu$ is imaginary or all $\nu$'s are irrelevant, and grey is when any of the conditions 1, 2 or 3 is violated.

\begin{figure}[]
\centering
\parbox{4.5cm}{
\includegraphics[scale=0.5]{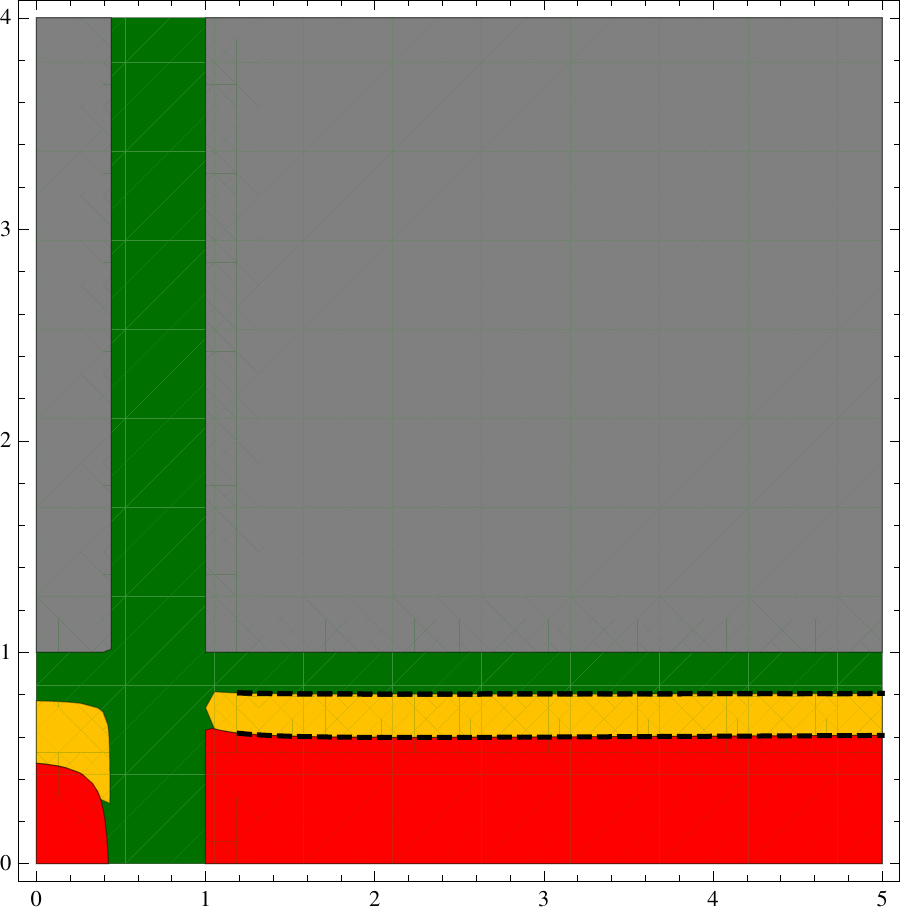}
\caption{Plot of $\eta$ v.s $\tilde{z}$ for $\gamma=2$ and $c_1=3$ \label{fig:13}  }
\label{fig:2figsA9}}
\qquad
~ ~ ~ ~ ~ ~ ~ ~ ~ ~ 
\begin{minipage}{4.5cm}
\includegraphics[scale=0.5]{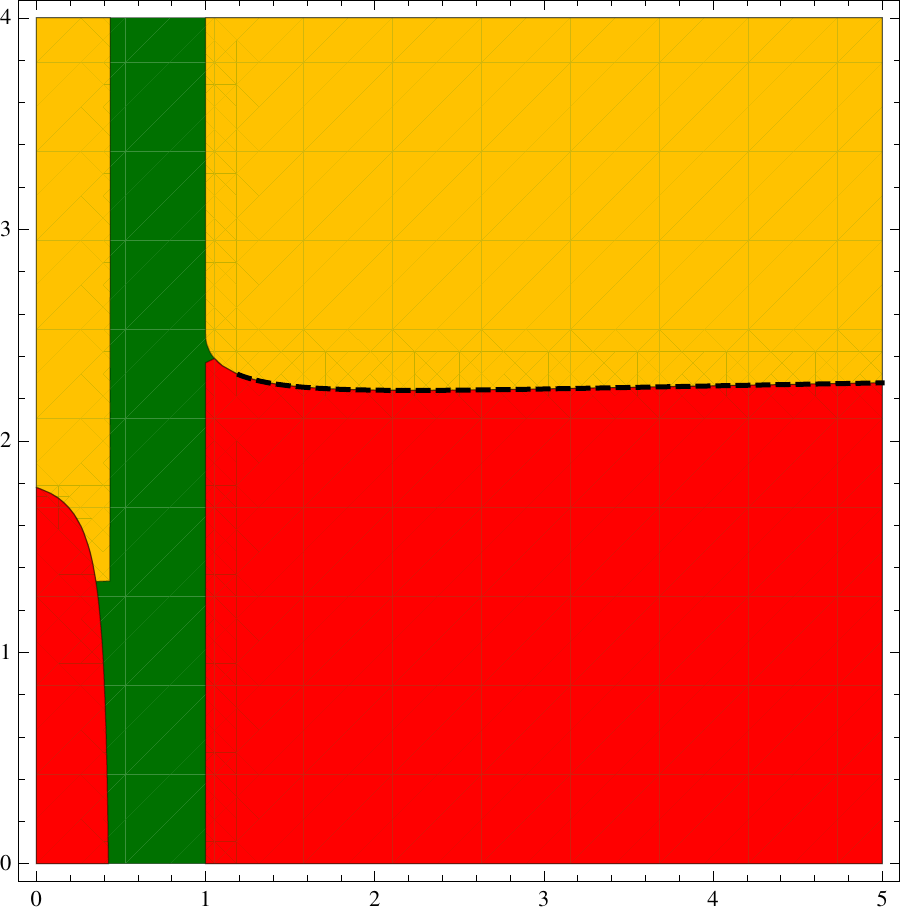}
\caption{Plot of $\eta$ v.s $\tilde{z}$ for $\gamma=2$ and $c_1=0.8$ \label{fig:14}     }
\label{fig:2figsB9}
\end{minipage}

\end{figure}

\begin{figure}[]
\centering
\parbox{4.5cm}{
\includegraphics[scale=0.5]{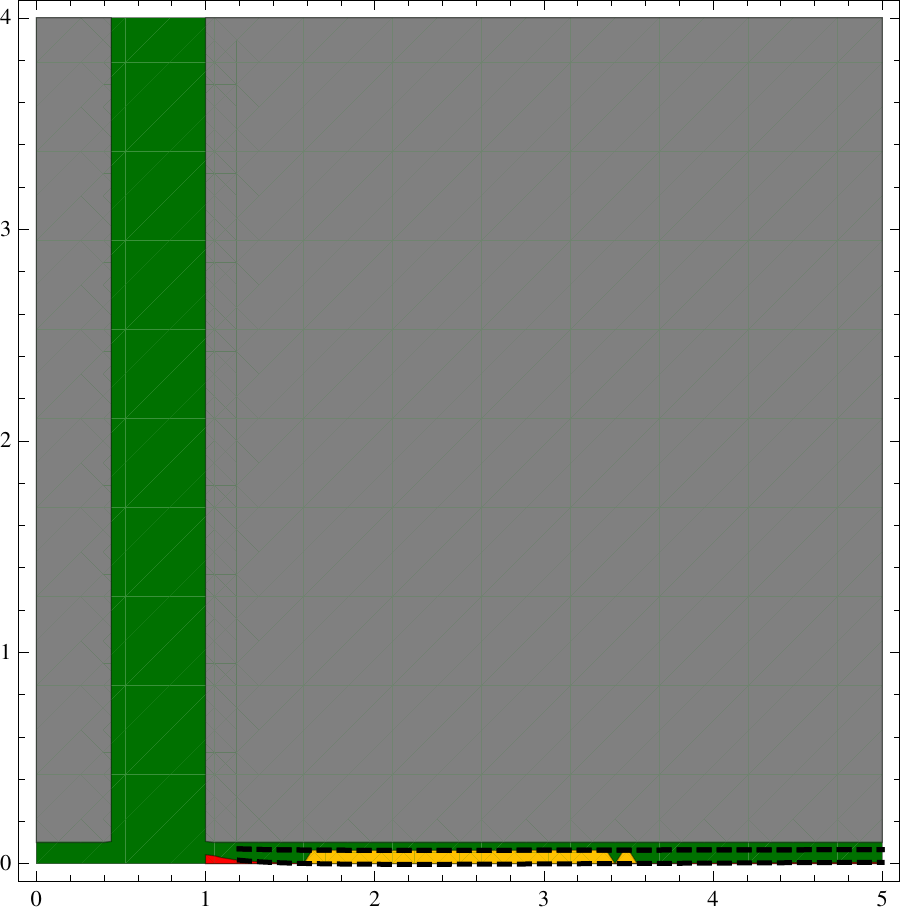}
\caption{Plot of $\eta$ v.s $\tilde{z}$ for $\gamma=0.2$ and $c_1=3$  \label{fig:15}  }
\label{fig:2figsA9}}
\qquad
~ ~ ~ ~ ~ ~ ~ ~ ~ ~ 
\begin{minipage}{4.5cm}
\includegraphics[scale=0.5]{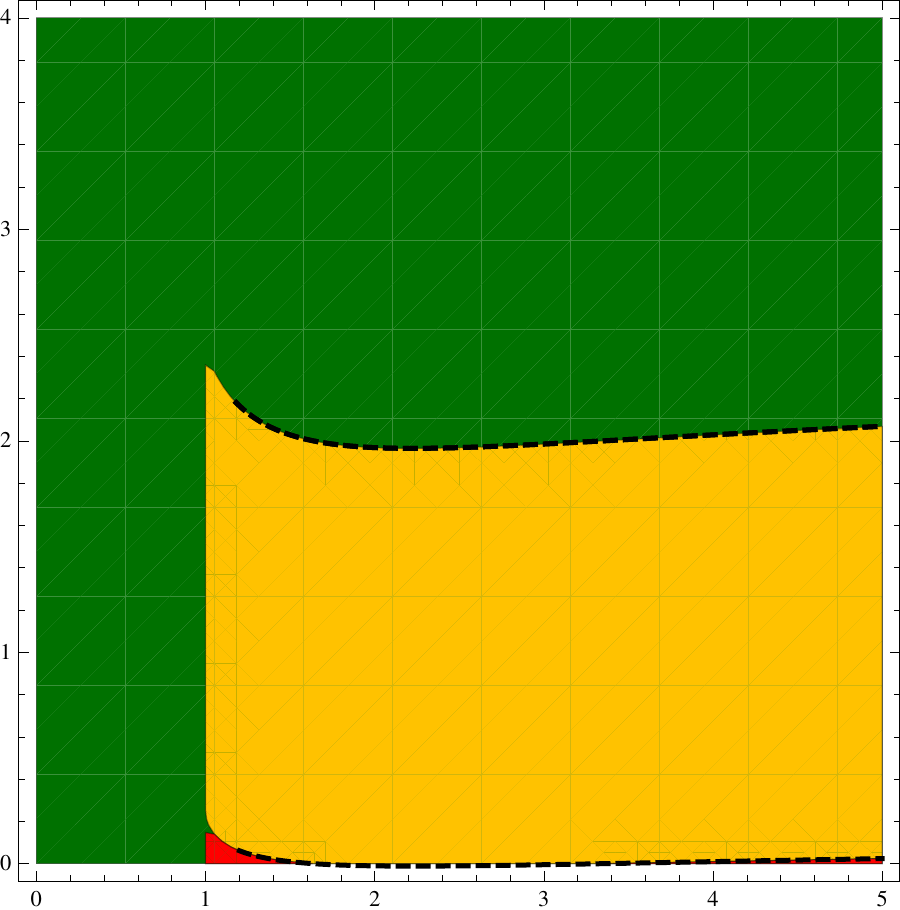}
\caption{Plot of $\eta$ v.s $\tilde{z}$ for $\gamma=0.2$ and $c_1=0.8$ \label{fig:16}    }
\label{fig:2figsB9}
\end{minipage}
\end{figure}

We found that for  $c_1< 1 $ and $\gamma >1$, there is no green region. But for $c_1> 1 $ for both cases of $\gamma >1$ and $\gamma <1$, green regions do exist. From figures~\ref{fig:13} -~\ref{fig:16}, one can notice that specifically the region of  $0< \tilde{z} <1$ is green, which indicates that $z>1$, consistent with casualty condition of HSV. So the green region is the most restricted region for the parameter space coming from the condition where the background is non-singular.\\

\begin{figure}[ht!] \label{Fig:figs}
\centering
\parbox{4.5cm}{
\includegraphics[scale=0.5]{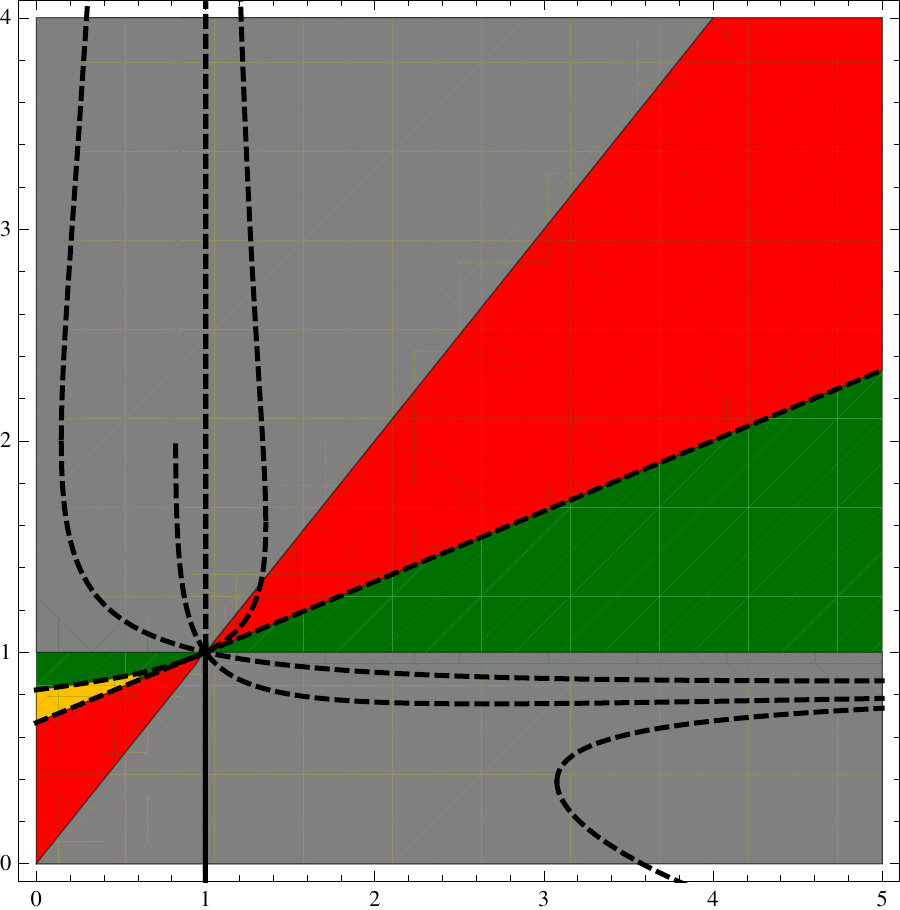}
\caption{Plot of $\eta$ v.s $\lambda$ for $\gamma=2$ and $c_1=3$ \label{fig:9} }
\label{fig:2figsA9}}
\qquad
~ ~ ~ ~ ~ ~ ~ ~ ~ ~ 
\begin{minipage}{4.5cm}
\includegraphics[scale=0.5]{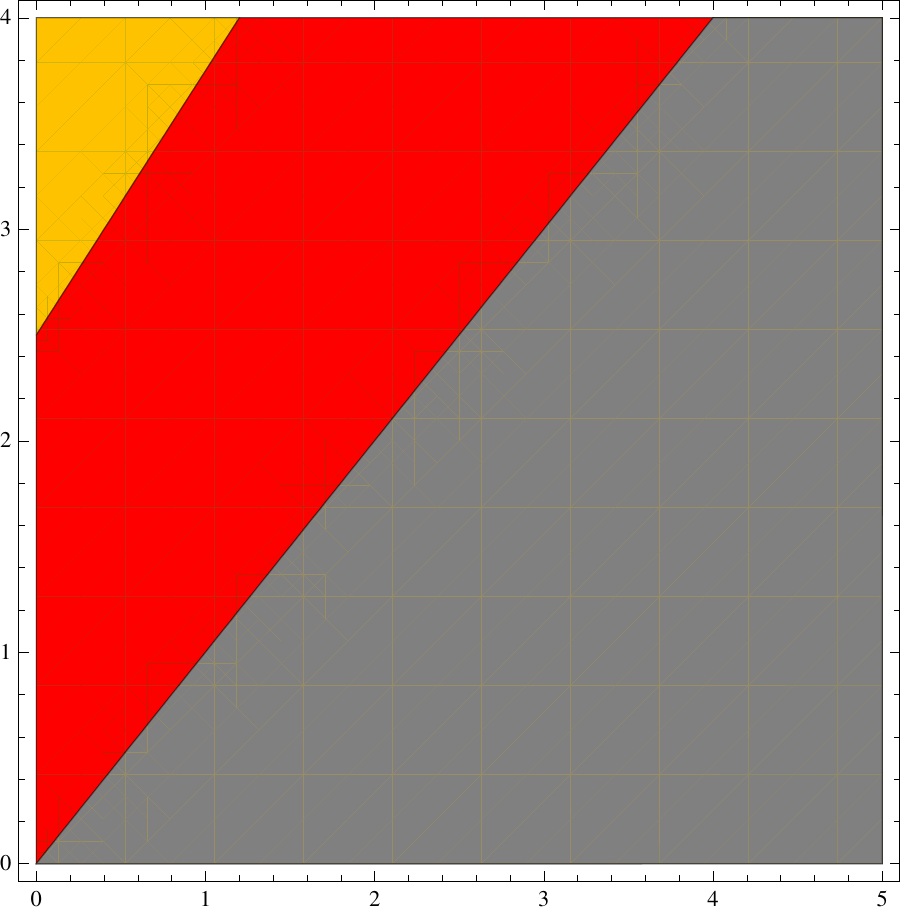}
\caption{Plot of $\eta$ v.s $\lambda$ for $\gamma=2$ and $c_1=0.8$  \label{fig:10}   }
\label{fig:2figsB9}
\end{minipage}

\centering
\parbox{4.5cm}{
\includegraphics[scale=0.5]{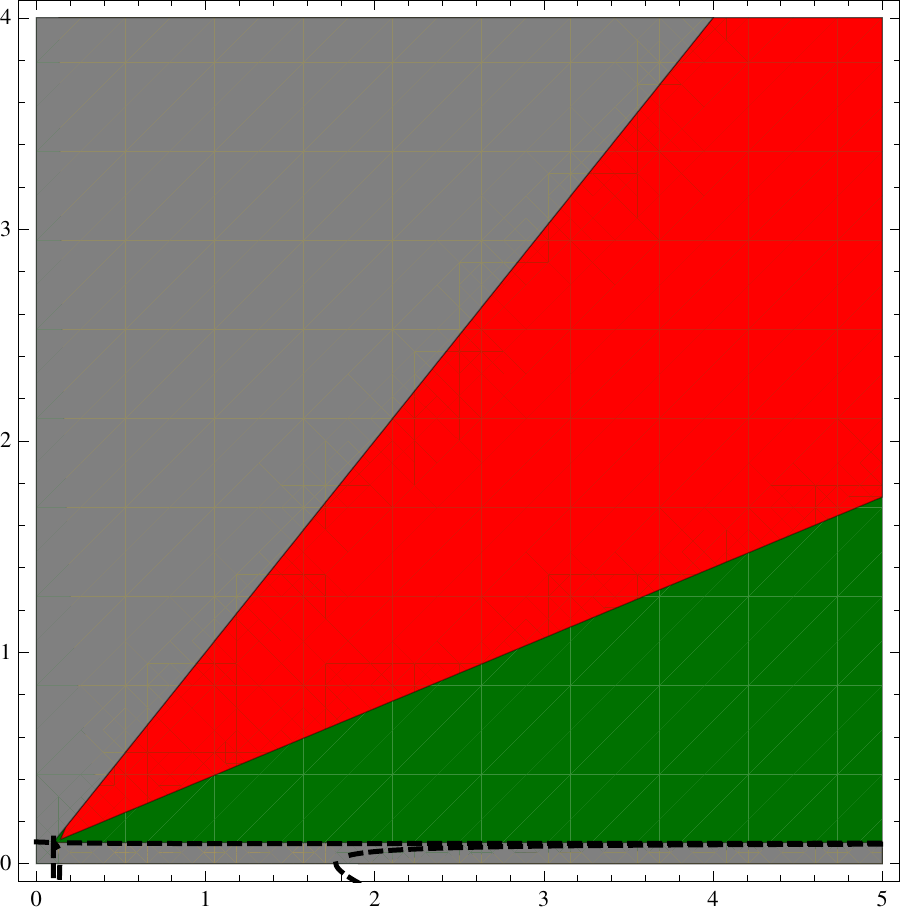}
\caption{Plot of $\eta$ v.s $\lambda$ for $\gamma=0.2$ and $c_1=3$ \label{fig:11}   }
\label{fig:2figsA9}}
\qquad
~ ~ ~ ~ ~ ~ ~ ~ ~ ~ 
\begin{minipage}{4.5cm}
\includegraphics[scale=0.5]{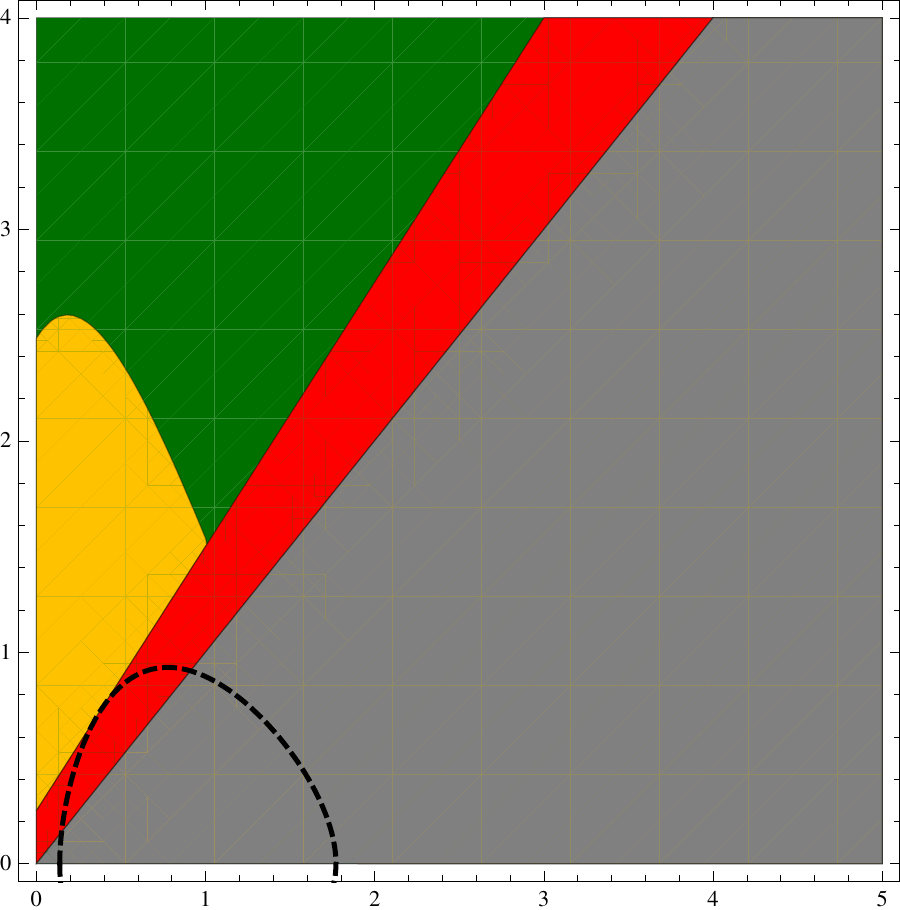}
\caption{Plot of $\eta$ v.s $\lambda$ for $\gamma=0.2$ and $c_1=0.8$  \label{fig:12}   }
\label{fig:2figsB9}
\end{minipage}

\end{figure}

For plotting $\eta$ versus $z$, we have used the following equation that we have derived in section \ref{ab},

\begin{equation}
\lambda =\frac{\theta -4}{\left(\frac{\text{   }L^2 (\theta -2)(-\theta +z (2+\theta ))+2\text{  }c_0z (\theta -2)(3+z (-4+\theta )-\theta )}{L^2 (z-1)}\right){}^{\frac{1}{2}}}.
\end{equation}

For plotting Figures~\ref{fig:9} -~\ref{fig:16}, we made the following assumptions
\begin{gather*}
\phi_0=0, \ \ \ L=1, \ \ \ c_0=-2, \ \ \  \theta=3.
\end{gather*}

\subsection{ Crossover Estimations}
We can analyze the physics of the flow by doing several estimations. The flow is from $\text{AdS}_4$ in the $\text{UV}$ to $\text{AdS}_2 \times \mathbb{R}^2$ in the $\text{IR}$ where each of these two regions has a constant dilaton. There is an intermediate hyperscaling violating region with solution \eqref{eq:newsolg} which it's dilaton flows logarithmically based on the relation $\phi(r)=\phi_0+ (\tilde{z}-1)\beta \log r $. One can approximately say that the exponential potential $V(\phi)$ is responsive for the HSV region, $f(\phi)\ F^2$ is responsible for the $\text{AdS}_4$, and $g(\phi)$ times the higher derivative terms is responsible for the emergence of $\text{AdS}_2 \times \mathbb{R}^2$ in the  $\text{IR}$. Using this we can estimate the $r$ and $\phi$ for each cross over. The cross over from $\text{AdS}_4$ in the UV to HSV happens at $\phi_U$ and $r_U$, and when $f(\phi_U)\ g^{rr} g^{tt} (F_{rt})^2 \sim V(\phi_U)$. Using this
\begin{gather}   
r_U=\Big(\frac{-1}{\tilde{L}^4} \ \frac{V_0 e^{\gamma \phi_0}}{\rho^2 e^{-\lambda \phi_0}}\Big)^{\frac{-1}{4(\theta(\tilde{z}-1)+2) }}, \nonumber\\
\phi_U=\phi_0-\frac{\beta (\tilde{z}-1)} {4 \theta(\tilde{z}-1)+8} \ \text{log}\Big(\frac{-V_0 e^{\gamma \phi_0}}{\tilde{L}^4  \ \rho^2 e^{-\lambda \phi_0}}\Big).
\end{gather}

For the first estimation, we take $\alpha_R=\alpha_W=\alpha_{GB}=L=1, \phi_0=0$, and $z=3, \ \theta=4$, leading to $\tilde{z}=\frac{2}{3}, \ \tilde{L}=0.693$. Also $c_0=-2$ and $c_1=3$, as in figures~\ref{fig:9},~\ref{fig:11}. This gives
\begin{gather}
r_U \simeq 0.32, \ \ \ \
\phi_U \simeq -6.45.
\end{gather}

After choosing the specific parameters, then $r_U$ and $\phi_U$ are fixed for the UV region. The crossover from HSV region to $\text{AdS}_2 \times \mathbb{R}^2$ in the $\text{IR}$ occurs at $\phi_I$ and $r_I$, when the higher derivative correction terms become comparable to the exponential potential. So $g(\phi_I)h \sim V(\phi_I)$, where 
\begin{gather}
h= \alpha_W C_{\mu\nu\rho\sigma} C^{\mu\nu\rho\sigma}+\alpha_{GB} G+\alpha_R R^2,
\nonumber
\end{gather}

and
\begin{gather}
g(\phi_I)= \frac{3}{4}(c_0 e^{\eta \phi_I}+c_1).
\end{gather}

For the assumed values we can calculate $h$, which gives, $h(r_I)=88 \ r_I^8$. This leads to the equation
\begin{gather}
r_I^8 (-2r_I^{\frac{4}{3}}+3)=-0.0292 r_I^{\frac{-4}{3}},
\end{gather}

which gives
\begin{gather}
r_I \simeq 1.356,  \ \ \ \
\phi_I \simeq 1.758.
\end{gather}

As $r_I$ is bigger than $r_U$, the cross over from $\text{AdS}_4$ to HSV happens after the cross over from HSV to $\text{AdS}_2 \times \mathbb{R}^2$ and so the RG flow can exist. Choosing a bigger $c_1$, which makes the effect of higher derivative terms more important and with all other parameters constant, one can easily make $r_I$ and $\phi_I$ bigger. As for example for $c_1=300$, we will get $r_I=43$. So one arbitrarily can increase the intermediate HSV region. The dilaton in $\text{AdS}_4$ and $\text{AdS}_2 \times \mathbb{R}^2$ is also constant with a bigger value in the UV. 

One can choose other parameter values and check whether for those cases, a flow could exist. There are some values that $r_U$ is bigger than $r_I$, or some other singularities can happen as indicated in the conditions for the existence of numerical flow in the previous section. For resolving those singularities, other methods should be implied. 

It would be much better to actually construct the numerical flow and see explicitly the interpolations between  $\text{AdS}_4$, intermediate hyperscaling violating region and $\text{AdS}_2 \times \mathbb{R}^2$. However due to the extensive parameters and the high sensitivity of the numerical solution to the initial values and parameters, a numerical flow could not be built explicitly here and can be done in future works. Shooting method similar to \cite{Knodel:2013fua} can be used to build the numerical flow, however for this case, it would be more difficult.

\section{Conclusions}

In this paper, first similar to \cite{Knodel:2013fua} we added squared curvature terms to the Einstein-Maxwell-dilaton action and replaced cosmological constant with a nontrivial dilatonic potential  to find a hyperscaling violating solution background. In order to set the $\alpha$ exponent of the ansatz with the exponent of exponential potential, we found out that we need to couple the higher derivative terms to the dilaton by a $g(\phi)$ function. After deriving the general analytical solution, because so many parameters play a role, we studied several different specific cases of the solution. We considered several constraints, null energy condition $\rho^2>0$, finiteness of the potential and the stability of the solution $\beta^2 >0$ and then with different combinations of $\alpha$ corrections we studied the behavior of the solution.

As the general hyperscaling violating solution similar to Lifshitz, has IR singularity where strings feel infinite tension there and therefore the solution is IR incomplete, we searched for a way to make the solution singularity free. Similar to \cite{Knodel:2013fua}, we attached the HSV solution to an $\text{AdS}_2 \times \mathbb{R}^2$ in the IR and $\text{AdS}_4$ in the UV by considering a flow from each region to the other. For doing so, we wrote the general solution in 4d in an (a-b) gauge which the form is similar in the all three regions, and then we studied the perturbations around $\text{AdS}_2 \times \mathbb{R}^2$ which can be used to shoot to HSV region. Then by implying several physical constraints, we derived allowed regions for the parameter values, and then by doing estimations for some specific parameters, we showed that the flow in theory can exist. Actually it would be nice to build the numerical flow using shooting techniques explicitly and then look for it's behavior in each region, which can be done in future works.

It is also worth to notice that we need to choose some specific values for the parameters in order to make such cross overs exist. For different other choices of parameters, different singularities which can be even stronger than HSV or Lifshitz singularity could occur, and some other methods need to be implied to resolve them.

The further study of this work can be as follows. The black brane solution of this kind can be constructed and then the thermodynamical properties of it can be studied. The dual field theory side of this general solution or the interpretation of the singularity and it's resolution  on the boundary can be studied. AC conductivity and the condition for the existence of zero sound and quasi-particle modes for this corrected gravity solution can also be calculated.

 Thermodynamics of hyperscaling violating background and the IR instabilities of the geometries has been studied in \cite{Hung}  \cite{Sadeghi}  \cite{Sinkovics}  \cite{Huijse}. Hyperscaling violating solution recently has been used to study thermalization \cite{Mozaffar14} \cite{Fonda}. It would also be useful to study thermalization and time dependent evolution and scaling behavior of entanglement entropy in this complete singularity free solution. However, there is no simple formula for calculating entanglement entropy with $\alpha$ corrections, except for the special case of $\alpha_{GB}$ \cite{Hung}, which can be calculated for our solution and then compared with the entanglement entropy in Einstein theory, in the future works.

 \section*{Acknowledgements}
I am very grateful to Mohammad Reza Mohammadi Mozaffar for extensive help during all the stages of this work specifically during finding the general solution, calculating the perturbations and also for many insightful discussions. I would like to thank Mohsen Alishahiha for initiating this work, reading the drafts and encouragements. I also would like to thank Shahin Sheikh Jabbari, James Liu,  Saedeh Sadeghian and Kamal Hajian for useful discussions. This work has been done at Institute for Research in Fundamental Sciences, IPM, Tehran, Iran.


\begin{thebibliography}{}


 
 
\bibitem{Maldacena}
J. M. Maldacena,
"The large N limit of superconformal field theories and supergravity,''
Adv.\ Theor.\ Math.\ Phys.\  {\bf 2}, 231 (1998)
[Int.\ J.\ Theor.\ Phys.\  {\bf 38}, 1113 (1999)]
[hep-th/9711200].


\bibitem{fixedpoints} 
  S. Kachru, X. Liu, M. Mulligan
  ``Gravity Duals of Lifshitz-like Fixed Points,''
  Phys. Rev. D78: 106005,2008, 
  [arXiv:0808.1725 [hep-th]].

\bibitem{Dong:2012se} 
  X.~Dong, S.~Harrison, S.~Kachru, G.~Torroba and H.~Wang,
  ``Aspects of holography for theories with hyperscaling violation,''
  JHEP {\bf 1206}, 041 (2012)
  [arXiv:1201.1905 [hep-th]].
\bibitem{0912.1061} 
S. A. Hartnoll, J. Polchinski, E. Silverstein, D. Tong,
 ``Towards strange metallic holography,''
   JHEP 1004:120,2010, [arXiv:0912.1061 [hep-th]].


\bibitem{Knodel:2013fua} 
  G.~Knodel and J.~T.~Liu,
 ``Higher derivative corrections to Lifshitz backgrounds,''
  [arXiv:1305.3279 [hep-th]].

\bibitem{Bap} 
 N. Bao, X. Dong, S. Harrison, E. Silverstein
  ``The Benefits of Stress: Resolution of the Lifshitz Singularity.''
  [arXiv:1207.0171 [hep-th]].

\bibitem{Carroll} 
 S. Carroll, ``Spacetime and Geometry, An Introduction to General Relativity,''
Adisson Weslay (2004).  

\bibitem{Shaghoulian:2013qia} 
  E.~Shaghoulian,
  ``FRW cosmologies and hyperscaling-violating geometries: higher curvature corrections, ultrametricity, Q-space/QFT duality, and a little string theory,''
  [arXiv:1308.1095 [hep-th]].


\bibitem{alishahiha:2012}
M.  Alishahiha, E. \'O Colg\'ain and  H. Yavartanoo, ``Charged Black Branes  with Hyperscaling Violating Factor,''    [arXiv:1209.3946 [hep-th]].


\bibitem{Sachdevbook} 
S. Sachdev. Quantum Phase Transitions " (2nd ed., Cambridge Univ. Press, 2010)
  
\bibitem{Senthil} 
 T. Senthil, S. Sachdev, M. Vojta,
  ``Fractionalized Fermi liquids''
   	Physical Review Letters 90, 216403 (2003) [arXiv:cond-mat/0209144].
  
\bibitem{Musso} 
 D. Musso
  ``Introductory notes on holographic superconductors''
    [arXiv:1401.1504 [hep-th]].
  
\bibitem{Sachdev2011de} 
 S. Sachdev,
  ``Holographic metals and the fractionalized Fermi liquid,''
  Physical Review Letters 105, 151602 (2010), [arXiv:1105.6335 [hep-th]].

  
\bibitem{Sachdev2014de} 
 A. Lucas, S. Sachdev, K. Schalm,
  ``Scale-invariant hyperscaling-violating holographic theories and the resistivity of strange metals with random-field disorder,''
  Phys. Rev. D 89, 066018 (2014) , [ arXiv:1401.7993 [hep-th]].
  
  
  
\bibitem{Roy} 
 P. Dey, S.Roy,
  ``Zero sound in strange metals with hyperscaling violation from holography,''
[arXiv:1307.0195 [hep-th]].
  
\bibitem{Mollabashi} 
 M. Reza Mohammadi Mozaffar, Ali Mollabashi
  ``Holographic quantum critical points in Lifshitz space-time,''
JHEP 04(2013)081, [arXiv:1212.6635 [hep-th]].
  
\bibitem{Ogawa11} 
 N. Ogawa, T. Takayanagi and T. Ugajin,
  ``Holographic Fermi Surfaces and Entanglement Entropy,''
  [arXiv:1111.1023 [hep-th]. 

  
\bibitem{Tarrio:2011de} 
  J.~Tarrio and S.~Vandoren,
  ``Black holes and black branes in Lifshitz spacetimes,''
  [arXiv:1105.6335 [hep-th]].

\bibitem{Goldstein:2010de} 
  K.Goldstein, Shamit Kachru, S.Prakash, P.Trivedi
  ``Holography of Charged Dilatonic Black Holes''
  JHEP {\bf 1109}, 017 (2011)
  [arXiv:0911.3586 [hep-th]].

\bibitem{Horowitz} 
Gary T. Horowitz, Benson Way
  ``Lifshitz Singularities''
  [arXiv:hep-th/1111.1243]

\bibitem{Lei} 
  Y. Lei, S. F. Ross
  ``Extending the nonsingular hyperscaling violeting space times''
  [arXiv:1310.5878 [hep-th]].
  
\bibitem{Cremonini} 
  J. Bhattacharya, S. Cremonini, A. Sinkovics
  ``On the IR completion of geometries with hyperscaling violation''
  [arXiv:1208.1752 [hep-th]].



\bibitem{Copsey} 
  K.Copsey, R.Mann
  ``Singularities in Hyperscaling Violation''
  [arXiv:1210.1231v2 [hep-th]].

\bibitem{Kachru} 
Sarah Harrison, Shamit Kachru and Huajia Wang
  ``Resolving Lifshitz Horizons''
  [arXiv:1202.6635[hep-th]].


\bibitem{Dadhich} 
Naresh Dadhich
  ``On the Gauss-Bonnet Gravity''
  [arXiv:hep-th/0509126]

\bibitem{Mozaffar} 
M. Alishahiha, M. R Mozaffar, A. Mollabashi
  ``Holographic Aspects of Two-charged Dilatonic Black Hole in AdS5 ''
  [arXiv: 1208.2535 ]

\bibitem{Shaghoulian} 
E. Shaghoulian
  ``Holographic Entanglement Entropy and Fermi Surfaces''
  [arXiv: 1112.2702]

\bibitem{Peet} 
D. K. O'Keeffe, A.W.Peet
  ``Electric hyperscaling violating solutions in
Einstein-Maxwell-dilaton gravity with R2 corrections''
  [arXiv: 1312.2261v1]
  
\bibitem{Mann} 
 K. Copsey and R. Mann
  ``Pathologies in Asymptotically Lifshitz Spacetimes''
 JHEP 1103, 039 (2011) [arXiv:1011.3502 [hep-th]].

  
\bibitem{Mozaffar14} 
M. Alishahiha, A. Faraji Astaneh, M. R. Mozaffar
  ``Thermalization in Backgrounds with Hyperscaling Violating Factor ''
 [arXiv:1401.2807[hep-th]].

  
  
\bibitem{Fonda} 
P. Fonda, Lasse Franti, V. Keranen, E. Keski-Vakkuri, L. Thorlacius, E. Tonni
  ``Holographic thermalization with Lifshitz scaling and hyperscaling violation ''
 [arXiv:1401.6088  [hep-th]].



  
\bibitem{Hung} 
L.Y. Hung, R. C. Myers and M. Smolkin
  ``On Holographic Entanglement Entropy and Higher Curvature Gravity''
 [arXiv:1101.5813 [hep-th]].

\bibitem{Iizuka} 
N. Iizuka, K. Maeda
  ``Stripe Instabilities of Geometries with Hyperscaling Violation''
 [arXiv:1301.5677 [hep-th]].

  
\bibitem{Sadeghi} 
J. Sadeghi, B. Pourhassan, A. Asadi
  ``Thermodynamics of string black hole with hyperscaling violation''
 [arXiv:1209.1235 [hep-th]].

  
  
\bibitem{Sinkovics} 
S. Cremonini, A. Sinkovics
  ``Spatially Modulated Instabilities of Geometries with Hyperscaling Violation''
 [arXiv:1212.4172 [hep-th]].

  \bibitem{Huijse} 
  L. Huijse, S. Sachdev and B. Swingle
  "Hidden Fermi surfaces in compressible states of gauge-gravity duality,"
  Phys. Rev. B 85, 035121 (2012) [arXiv:1112.0573 [cond-mat.str-el]].
  
  
  \bibitem{Goutéraux} 
  C. Charmousis, B. Goutéraux, B. S. Kim , E. Kiritsis, Rene Meyer  "Effective Holographic Theories for low-temperature condensed matter systems,"
  JHEP 1011:151,2010  [arXiv:1005.4690].
  
  \bibitem{Pang} 
 Rong-Gen Cai, Chiang-Mei Chen, Kei-ichi Maeda, Nobuyoshi Ohta, Da-Wei Pang "Entropy Function and Universality of Entropy-Area Relation for Small Black Holes,"
 Phys.Rev.D77:064030,2008  [arXiv:0712.4212].
  
  \bibitem{Pedraza} 
 M. Edalati, J. F. Pedraza "Aspects of Current Correlators in Holographic Theories with Hyperscaling Violation,"
Phys. Rev. D 88, 086004 (2013)  [arXiv:1307.0808].
  
  \bibitem{Odintsov} 
 E. Elizalde, A.G. Jacksenaev, S.D. Odintsov, I.L. Shapiro, "A Four-Dimensional Theory for Quantum Gravity with Conformal and Nonconformal Explicit Solutions,"
Class.Quant.Grav.12:1385-1400,1995 [arXiv:hep-th/9412061]. 
\end{thebibliography}
\end{document}